\documentclass[sts]{imsart}

\RequirePackage[OT1]{fontenc}
\usepackage{amsthm,natbib}
\RequirePackage[colorlinks,citecolor=blue,urlcolor=blue]{hyperref}

\usepackage{tkz-graph}  
\usepackage{amsmath}
\usepackage{color}
\usepackage{array}
\usepackage{rotfloat}
\usepackage{float}
\usepackage{multirow}
\usepackage{amsmath}
\usepackage{amssymb}
\usepackage{esint}
\usepackage{color}
\usepackage{array}
\usepackage{rotfloat}
\usepackage{multirow}
\usepackage{algorithm2e}
\usepackage{amssymb}
\usepackage{graphicx}
\usepackage{esint}

\startlocaldefs
\numberwithin{equation}{section}
\theoremstyle{plain}

\endlocaldefs

\providecommand{\tabularnewline}{\\}

\newcommand{\biblist}{\begin{list}{}
{\listparindent 0.0cm \leftmargin 0.50cm \itemindent -0.50 cm
\labelwidth 0 cm \labelsep 0.50 cm
\usecounter{list}}\clubpenalty4000\widowpenalty4000}
\newcommand{\ebiblist}{\end{list}}

\usepackage{latexsym}

\newtheorem{example}{Example}

\setcitestyle{authoryear,open={(},close={)}}

\usepackage[english]{babel}
\begin{document}

\begin{frontmatter}
\title{\LARGE Fractional Imputation in Survey
Sampling: A Comparative Review}
\runtitle{Fractional Imputation}
%\thankstext{T1}{Footnote to the title with the `thankstext' command.}

\begin{aug}
\author{\fnms{Shu} \snm{Yang}\ead[label=e1]{shuyang@hsph.harvard.com}} 
\and
\author{\fnms{Jae Kwang} \snm{Kim}\ead[label=e2]{jkim@iastate.edu}}
%\and
%\author{\fnms{Third} \snm{Author}\thanksref{t1}
%\ead[label=e3]{third@somewhere.com}
%\ead[label=u1,url]{www.foo.com}}

%\thankstext{t1}{Some comment}
%\thankstext{t2}{First supporter of the project}
\runauthor{S. Yang and J. K. Kim}

\affiliation{Harvard University and Iowa State University}

\address{Room 437A, HSPH2, 655 Huntington Ave, Boston, MA 02115
\printead{e1}.}

\address{
1208 Snedecor Hall, Iowa State University, Ames, IA 50011
\printead{e2}.}

%\address{Address of the Third author,
%usually few lines long,
%usually few lines long \printead{e3,u1}.}
\end{aug}

\begin{abstract}
Fractional imputation (FI) is a relatively new method of imputation for  handling item nonresponse in survey sampling. In FI, several imputed values with their fractional weights are created 
for each missing item. Each  fractional weight represents the conditional probability  of  the  imputed value given the observed data, and the parameters in the conditional probabilities are often computed by an iterative method such as EM algorithm. The underlying model for FI can be fully parametric, semiparametric, or nonparametric, depending on plausibility of assumptions and the data structure. 

In this paper, we give an overview of FI, 
introduce key ideas and methods to readers who are new to the FI literature, and highlight some new development. We also provide guidance on practical
implementation of FI and valid inferential tools after imputation. We demonstrate the empirical performance of FI with respect to multiple imputation using a pseudo finite population generated from a sample in Monthly Retail Trade Survey in US Census Bureau.

\end{abstract}


\end{frontmatter}

% citealt-citealt, citealt*-citealt*, citealt-citealt

\section{INTRODUCTION}

In survey sampling, it is a common practice to collect data on a large
number of items. Even when a sampled unit responds to the survey,
this unit may not respond to some items.
In this scenario, imputation can be used to create a complete data set by filling in missing values with plausible values to facilitate data analyses. The goal of imputation is three-fold: 
First, by providing  complete data, subsequent analyses are easy to implement and can achieve consistency among different users. 
Second, imputation reduces the selection bias associated with only using the respondent set, which may not necessarily represent the original sample.  
Third,  the imputed data can incorporate extra information so that the resulting analyses are statistically efficient and coherent.
Combining information from several surveys or creating synthetic data from planned missingness are cases in point (\citealt{schenker07}). 

When the imputed data set is released to the public, it should  
meet the goal of multiple uses both for planned and unplanned parameters
\citep{haziza2009imputation}. In a typical survey situation, the imputers may know some of the parameters of interest at the time of imputation, but hardly know the full set of possible parameters to be estimated from the data. 
Single imputation, such as hot deck imputation, regression imputation and
stochastic regression imputation, replaces each of the missing data
with one plausible value. Although
single imputation has been widely used, one drawback is that it does
not take into account of the full uncertainty of missing data and often falls short of  multiple-purpose estimation.
Multiple imputation (MI) has been proposed by \citet*{rubin1976inference}
to replace each of missing data with multiple plausible values to reflect the full uncertainty in the prediction of missing data. 
%MI has since then become a popular  approach of imputation for general-purpose estimation. 
Several authors (\citealt{rubin1987multiple}; \citealt{little2002statistical}; \citealt{schafer1997imputation}) have promoted MI as a standard approach for general-purpose estimation under item nonresponse in survey sampling.
Although the variance estimation formula of \citet*{rubin1987multiple}
is simple and easy to apply, it is not always consistent (\citealt{fay1992inferences};
\citealt{wang1998large}; \citealt{kim2006bias}). For using the MI variance estimation formula, the congeniality condition
of \citet*{meng1994multiple} needs to be met, which can be restrictive
for general-purpose inference. For example, \citet*{kim11}
pointed out that a 
 MI procedure that is 
  congenial for mean estimation is
not necessarily congenial for proportion estimation. 

Fractional imputation (FI) is another effective imputation tool for general-purpose estimation 
with its advantage of not requiring the congeniality condition. FI
was originally proposed by \citet*{kalton1984some} to reduce the variance
of single imputation methods by replacing each missing value with
several plausible values at differentiable probabilities reflected through fractional weights. \citet*{fay1996alternative},
\citet*{kim2004fractional}, \citet*{fuller2005hot}, \citet*{durrant2005imputation},
\citet*{durrant2006using} discussed FI as a nonparametric imputation
method for descriptive parameters of interest in survey sampling.  \citet*{kim11} and \citet*{Kim2014Fractionalhotdeck} presented FI under fully parametric model assumptions.

More generally, FI can also serve as a computational tool
for implementing the  expectation step (E-step) in the EM algorithm (\citealt{wei1990monte};
\citealt{kim11}).  When the conditional expectation in the E-step is 
not available in a closed form, parametric FI of \citet*{kim11}
simplifies computation by drawing on the importance
sampling idea. Through fractional weights, FI can reduce the burden of iterative
computation, such as Markov Chain Monte Carlo, for evaluating
the conditional expectation associated with missing data. 
\citet*{kimhong12} extended parametric FI to a more general class of incomplete data, including measurement error models.

%The advantages of this FI approach lie in its statistical validity and computational
%simplicity. 

Despite these advantages, 
%the application of FI in applied research
%has been relatively infrequent. 
FI in applied research has not been widely used due to lack of good 
information that provides researchers with comprehensive understanding 
of this approach. 
The goal of this paper is to bring more attention to FI by reviewing existing research on FI, introducing key ideas and methods, and highlighting some new development, mainly
in the context of survey sampling. This paper also provides guidance
on practical implementations and applications of FI. 

This paper is organized as follows. Section 2 provides the basic
setup and Section 3 introduces FI under parametric model assumptions.
Section 4 discusses a nonparametric approach to FI, specially in the context of hot deck imputation. Section 5 introduces synthetic data imputation using FI in the context of two-phase sampling and statistical matching. 
Section 6 deals with practical considerations and variations of FI, including imputation sizes, choices of proposal distributions and doubly robust FI.
Section 7 compares FI with MI in terms of efficiency of the point
estimator and the variance estimator. Section 8 demonstrates a
simulation study based on an actual data set. A discussion concludes
this paper in Section 9.

\section{BASIC SETUP}

Consider a finite population of $N$ units identified by a set of
indices $U=\{1,2,\cdots,N\}$ with $N$ known. 
The $p$-dimensional study variable $y_{i}=(y_{i1},\cdots,y_{ip})$, associated with each
unit $i$ in the population,
is subject to missingness. We assume that the finite population at hand is a realization from an infinite population, called a \textit{superpopulation}. In the
superpopulation model, we often postulate a parametric distribution,
$f(y;\theta)$, with the parameter $\theta\in\Omega$. We can express the density for the joint distribution of $y$ as 
\begin{equation}
f(y ; \theta)=f_{1}(y_{1} ;\theta_{1})f_{2}(y_{2}\mid y_{1};\theta_{2})\cdots f_{p}(y_{p}\mid y_{1},\cdots,y_{p-1};\theta_{p}) \label{eq: joint dist}
\end{equation}
where $\theta_{k}$ is the parameter in the conditional distribution
of $y_{k}$ given $y_{1},\cdots,y_{k-1}$. 
Now let $A$ denote the set of indices for units in a sample selected
by a probability sampling mechanism. Each unit
is associated with a sampling weight, the inverse of the probability
of being selected  to the sample, denoted by $w_{i}$. 
%In the non-survey sampling setups, we consider $w_{i}$ to be equal for all units. 

We are interested in  estimating $\eta$, defined as
a (unique) solution to the population estimating equation $\sum_{i=1}^{N}U(\eta;y_{i})=0$.
For example, a population mean of $y$ can be obtained by letting
$U(\eta;y_{i})=\eta-y_{i}$, a population proportion of $y$ less
than a threshold $c$ can be obtained by specifying $U(\eta;y_{i})=\eta-I_{\{y_{i}<c\}}$,
where $I$ is an indicator function, a population median of $y$
can be obtained by choosing $U(\eta;y_{i})=0.5-I_{\{y_{i}<\eta\}}$, and so on.
Under complete response, a consistent estimator of $\eta$ is obtained
by solving 
\begin{equation}
\sum_{i\in A}w_{i}U(\eta;y_{i})=0.\label{eq:compEE}
\end{equation}
\citet*{godambe1986parameters}, \citet*{binder1994use} and \citet*{rao2002estimating} have done rigorous investigations on the estimator obtained from (\ref{eq:compEE}) under complex sampling.

%In the presence of missing data, 
In the presence of missing data, first consider decomposing $y_{i}=(y_{obs,i},y_{mis,i})$, where 
$y_{obs,i}$ and $y_{mis,i}$ are the observed and missing
part of $y_{i}$, respectively. 
%Let $\delta_i$ be the indicator of full response, that is, if $\delta_i=1$, we have $y_{obs,i}=y_i$.
We assume that the response mechanism
is missing at random (MAR) in the sense of \citet*{rubin1976inference}. 
That is, the probability of nonresponse does not depend on the missing
value itself. Under MAR, 
a consistent estimator of $\eta$
can be obtained by solving the conditional estimating equation, given the observed data $y_{obs}=(y_{obs,1},\ldots,y_{obs,n})$, 
\begin{equation}
\sum_{i\in A}w_{i}E\{U(\eta;y_{i})\mid y_{obs,i}\}=0,\label{eq:condEE}
\end{equation}
where the above conditional expectation  is taken with respect to the prediction model (also called the imputation model), 
\begin{equation}
 f( y_{mis, i} \mid  y_{obs, i}; \theta )  = \frac{ f( y_{obs,i}, y_{mis, i} ;  \theta )} { \int  f( y_{obs,i}, y_{mis, i} ; \theta ) d y_{mis, i} } ,
\label{4}
\end{equation}
which depends on the unknown parameter $\theta$. Imputation is thus a computational tool for computing  the conditional expectation in (\ref{eq:condEE}) for arbitrary choices of the estimating function $U(\eta; y)$.  The resulting conditional expectation using imputation can be called the imputed estimating function. 

Table 1 presents a summary of Bayesian and frequentist approaches of statistical  inference with missing data. In the Bayesian approach, $\theta$ is treated as a random variable and the reference distribution is the joint distribution of $\theta$ and the latent (missing) data, given the observed data. On the other hand, in the frequentist approach, $\theta$ is treated as fixed and the reference distribution is the conditional distribution of the latent data, conditional on the observed data, for a given parameter $\theta$. The learning algorithm, that is, the algorithm for updating information for parameters from observed data,  for the Bayesian approach is data augmentation (\citealt{Tanner87}), while the learning algorithm for the frequentist approach is usually the EM algorithm.

\begin{table}[t]
  \begin{center}
  \caption{Comparison of two approaches of inference with missing data} 
  \begin{tabular}{ccc}
      \hline%
    & Bayesian & Frequentist \\
    \hline 
    Model & Posterior distribution & Prediction model \\
     & $f( \mbox{latent}, \theta \mid \mbox{Obs.} )$ & $f( \mbox{latent} \mid \mbox{Obs.}, \theta ) $\\
    \hline 
  Learning algorithm& Data augmentation & EM algorithm \\
   Prediction  & Imputation(I)-step & Expectation(E)-step \\
   Parameter update & Posterior(P)-step & Maximization(M)-step \\
  \hline
  Imputation & Multiple imputation & Fractional imputation \\
  \hline 
  Variance estimation & Rubin's formula & Linearization \\ 
  & & or replication \\
  \hline 
\end{tabular}
  \end{center}
\end{table}

MI is a Bayesian imputation method and  
 the imputed estimating function is computed with respect to  
 the posterior predictive distribution,
$$ f( y_{mis, i} \mid  y_{obs} )  = \int f( y_{mis, i} \mid  y_{obs, i} ; \theta)  p( \theta \mid y_{obs} )  d \theta,
$$
which is the average of  the predictive distribution $ f( y_{mis, i} \mid  y_{obs, i}; \theta ) $  over the posterior distribution of $\theta$.  
On the other hand, in the frequentist approach, the conditional expectation in (\ref{eq:condEE}) is taken with respect to the prediction model (\ref{4}) evaluated at $\theta=\hat{\theta}$, a consistent estimator of $\theta$. For example, 
one can use the pseudo MLE $\hat{\theta}$ of $\theta$ obtained
by solving the pseudo mean score equation (\citealt{louis82}; 
\citealt{pfeffermann1998weighting}),
\begin{equation}
\bar{S}(\theta)=\sum_{i\in A}w_{i}E\{S( {\theta}; y_i )|  y_{i,obs}; \theta \}=0,\label{eq:mean-score}
\end{equation}
where $S(\theta; y_i)=\partial\log f(y_{i} ; \theta)/\partial\theta$.

While the Bayesian approach to imputation, especially  in the context of MI, is well studied in the literature, the frequentist approach to imputation is somewhat sparse.   FI has been proposed to fill in this important gap.  In FI, the conditional expectation in (\ref{eq:condEE}) is computed by a weighted mean of the imputed estimating functions 
\begin{equation}
E\{U(\eta; y_{i}) \mid y_{obs,i}\}\cong\sum_{j=1}^{M}w_{ij}^{*}U(\eta; y_{obs,i},y_{mis,i}^{*(j)}).%\label{eq:approximation2}
\label{1-3}
\end{equation}
where 
$y_{mis,i}^{*(j)}$, for $j=1,\ldots,M$, are $M$ imputed values for $y_{mis,i}$
(if ${y}_{i}$ is completely observed, $y_{mis,i}^{*(j)}\equiv y_{mis,i}$),
$w_{ij}^*$ are the fractional weights that satisfies $w_{ij}^* \ge 0$, $\sum_{j=1}^M w_{ij}^* = 1$ and 
$$ \sum_{i\in A}w_{i}\sum_{j=1}^{M}w_{ij}^{*}S(\hat{\theta} ; y_{obs,i},y_{mis,i}^{*(j)})=0.$$
%Creating the imputed values and their fractional weights will be discussed in Section 3. 
Once the FI data are constructed, the FI estimator of $\eta$ is obtained by solving 
\begin{equation}
\sum_{i\in A}w_{i}\sum_{j=1}^{M}w_{ij}^{*}U(\eta; y_{obs,i},y_{mis,i}^{*(j)})=0.
\label{1-5}
\end{equation}

In general,  the FI method augments the original data set as
\begin{equation}
\mathcal{S}_{FI}= \left\{ \delta_i \left( w_i, y_i \right)+ (1-\delta_i) \left( w_iw_{ij}^*,  y_{ij}^* \right); j=1,\ldots, M, i \in A  \right\},
\label{1-6}
\end{equation}
where $\delta_i$ is the indicator of full response for $y_i$, and $y_{ij}^*=(y_{obs,i},y_{mis,i}^{*(j)})$.
If (\ref{1-3}) holds for an arbitrary $U$ function, the resulting estimator is approximately unbiased for a fairly large class of parameters, 
  which makes the imputation attractive for general-purpose estimation.  \citet*{kim11} used the importance sampling technique to satisfy (\ref{1-3}) for general $U$ functions, which will be presented in the next section.

\section{PARAMETRIC FRACTIONAL IMPUTATION}
\label{sec:PFI}

\textit{Parametric Fractional Imputation} (PFI),  
proposed by \citet*{kim11}, features a parametric model for fractional imputations, and parameters in the imputation model are estimated by a computationally efficient EM algorithm. %where the imputed values are generated by the importance sampling at the beginning of the iteration and the fractional weights are updated for each EM iteration. 

To  compute the conditional estimating equation in (\ref{eq:condEE}) by PFI, for each missing value $y_{mis,i}$, generate $M$ imputed values, denoted by $\{y_{mis,i}^{*(1)},\ldots,y_{mis,i}^{*(M)}\}$
from a proposal distribution $h(y_{mis,i}\mid y_{obs,i})$.  How to choose a
proposal distribution will be discussed in Section \ref{sub:Choice of h}.
Once the imputed values are generated from $h(\cdot)$, compute 
\[
w_{ij}^{*}\propto\frac{f(y_{mis,i}^{*(j)}\mid y_{obs,i};\hat{\theta})}{h(y_{mis,i}^{*(j)}\mid y_{obs,i})},
\]
subject to $\sum_{j=1}^{M}w_{ij}^{*}=1$, as the fractional weights  assigned to
$y_{ij}^* = (y_{obs,i},y_{mis,i}^{*(j)})$, where $\hat{\theta}$
is the pseudo MLE of $\theta$ to be determined by the EM algorithm below.
Since $\sum_{j=1}^{M}w_{ij}^{*}=1$, the above fractional weight is
the same as $w_{ij}^* = w_{ij}^{*} ( \hat{\theta})$, where  
\begin{equation}
 w_{ij}^{*} ( {\theta}) \propto\frac{f(y_{obs,i},y_{mis,i}^{*(j)}; {\theta})}{h(y_{mis,i}^{*(j)}\mid y_{obs,i})},
\label{9}
\end{equation}
which only requires the  knowledge of the joint distribution $f(y ;\theta)$
and the proposal distribution $h$. 

The pseudo MLE of $\theta$ can be computed by solving the imputed mean score equation,
\begin{equation}
\sum_{i\in A} w_i \sum_{j=1}^{M}w_{ij}^{*} (\theta) S({\theta};y_{obs,i},y_{mis,i}^{*(j)})=0.\label{eq:approximation1}
\end{equation}
To solve (\ref{eq:approximation1}), we can either use the Newton method or 
the  following EM algorithm:
\begin{description}
\item [{\textit{{I-step.}}}] For each missing value $y_{mis,i}$, $M$ imputed
values are generated from a proposal distribution $h(y_{mis,i}\mid y_{obs,i})$. 
\item [{\textit{{W-step.}}}] Using the current value of the parameter
estimates $\hat{\theta}_{(t)}$, compute the fractional weights as
$w_{ij(t)}^{*}\propto f(y_{obs,i},y_{mis,i}^{*(j)} ;\hat{\theta}_{(t)})/h(y_{mis,i}^{*(j)}\mid y_{obs,i})$,
subject to $\sum_{j=1}^{M}w_{ij(t)}^{*}=1$. 
\item [{\textit{{M-step.}}}] Update the parameter $\hat{\theta}_{(t+1)}$
by solving the imputed score equation, $$\sum_{i\in A}w_{i}\sum_{j=1}^{M}w_{ij(t)}^{*}S(\theta; y_{ij}^{*})=0,$$
where $y_{ij}^{*}=(y_{obs,i},y_{mis,i}^{*(j)})$ and $S(\theta; y)=\partial\log f(y ;\theta)/\partial\theta$
is the score function of $\theta$.
\item [\textit{{Iteration.}}] Set $t=t+1$ and go to the W-step. Stop if $\hat{\theta}_{(t+1)}$
meets the convergence criterion. 
%\item [{\textit{{C-step.}}}] (Optional) Check if $w_{ij(t)}^{*}>  3 M^{-1}$
%for some $j=1,\ldots,M$. If yes, update the proposal distribution
%with $\hat{\theta}_{(0)}$ replaced by $\hat{\theta}_{(t)}$ and go
%to I-step. If no, go to M-step. Stop if $\hat{\theta}_{(t+1)}$
%meets some user-defined convergence criterion. 
\end{description}
Here, the I-step is the imputation step, the W-step is the weighting step, and the M-step
is the maximization step. The I- and W-steps can be
combined to implement the E-step of the EM algorithm. Unlike the Monte Carlo EM (MCEM)
method, imputed values are not changed for each EM iteration -- only
the fractional weights are changed. Thus, the FI method has computational
advantages over the MCEM method. Convergence is achieved because the
imputed values are not changed. \citet*{kim11} showed
that given the $M$ imputed values, $y_{mis,i}^{*(1)},\ldots,y_{mis,i}^{*(M)}$,
the sequence of estimators $\{\hat{\theta}_{(0)},\hat{\theta}_{(1)},\ldots\}$
from the W-and M-steps converges to a stationary point
$\hat{\theta}_{M}^{*}$ for fixed $M$. The stationary point $\hat{\theta}_{M}^{*}$
converges to the pseudo MLE of $\theta$ as $M\rightarrow\infty$.
The resulting weight $w_{ij}^{*}$ after convergence is the fractional
weight assigned to $y_{ij}^{*}=(y_{obs,i},y_{mis,i}^{*(j)})$.
%The C-step is used to assess the distribution of the resulting fractional weights.
%If several extremely large  weights dominate the weights,
%it indicates that the proposal distribution is not well-specified. A
%simple update is to use the proposal distribution with $\hat{\theta}_{(0)}$
%replaced by $\hat{\theta}_{(t)}$ and go to the $I$-step. 
We may add an additinal step to monitor the distribution of the fractional weights so that 
no extremely large fractional weights dominate the weights. 

Once the fractional imputed data is constructed from the above steps, it can be used to estimate other parameters of interest. That is, we can use (\ref{1-5}) to estimate $\eta$ from the FI data set. 

We now consider a bivariate missing data example to illustrate
the use of the EM algorithm in FI. 
\begin{example} 
Suppose a probability sample consists
of $n$ units of $z_i = (x_{i},y_{1i},y_{2i})$ with sampling weight $w_{i}$,
where  $x_i$ is always observed and 
$y_{i}=(y_{1i},y_{2i})$ is subject to missingness.  Let $A_{11}$, $A_{10}$, $A_{01}$, and $A_{00}$
be the partition of the sample based on the missing pattern, where
subscript $1$/$0$ in the $i$-th position denote that the $i$-th
$y$ item is observed/missing, respectively.  For example, $A_{10}$ is the set of
the sample with $y_{i1}$ observed and $y_{i2}$ missing.

The conditional expectation in (\ref{eq:condEE}) involves evaluating
the conditional distribution of $y_{mis,i}$ given the observed data
$x_{i}$ and $y_{obs,i}$ for each missing pattern, which is then decomposed into 
\begin{multline*}
\sum_{i\in A}w_{i}\mathbb{E}\{U(\eta;z_{i})|x_{i},y_{obs,i}\}=\sum_{i\in A_{11}}w_{i}U(\eta;x_{i},y_{i1},y_{i2})+\sum_{i\in A_{00}}w_{i}E\{U(\eta;x_{i},Y_{i1}, Y_{i2})\\ \quad \ \mid x_{i}\}
+\sum_{i\in A_{01}}w_{i}E\{U(\eta;x_{i},Y_{i1},y_{i2})\mid x_{i},y_{i2}\}+\sum_{i\in A_{10}}w_{i}E\{U(\eta;x_{i},y_{i1},Y_{i2})\mid x_{i},y_{i1}\}.
\end{multline*}
Suppose the joint distribution in (\ref{eq: joint dist})   is
\begin{equation}
f(x, y_{1},y_{2};\theta)= f_x ( x; \theta_0) f_{1}(y_{1}\mid x;\theta_{1})f_{2}(y_{2}\mid x,y_{1};\theta_{2}).\label{eq:jointDistp=00003D2}
\end{equation}
From the full respondent sample in $A_{11}$, obtain $\hat{\theta}_{1(0)}$
and $\hat{\theta}_{2(0)}$, which are initial parameter estimates
for $\theta_{1}$ and $\theta_{2}$. 

In the I-step, for each missing value $y_{mis,i}$, generate $M$
imputed values from $h(y_{mis,i}\mid x_{i},y_{obs,i})=f(y_{mis,i}\mid x_{i},y_{obs,i};\hat{\theta}_{(0)})$,
where 
\begin{equation}
f(y_{mis,i}\mid x_{i},y_{obs,i};\hat{\theta}_{(0)})=\left\{ \begin{array}{ll}
f_{2}(y_{2i}\mid x,y_{1i};\hat{\theta}_{2(0)}) & \mbox{ if }i\in A_{10}\\
f(y_{1i}\mid x,y_{2i};\hat{\theta}_{(0)}) & \mbox{ if }i\in A_{01}\\
f(y_{1i},y_{2i}\mid x_{i};\hat{\theta}_{(0)}) & \mbox{ if }i\in A_{00}
\end{array}\right.\label{6-8}
\end{equation}
and 
\begin{equation}
f(y_{1i}\mid x_{i},y_{2i};\hat{\theta}_{(0)})=\frac{f_{1}(y_{1i}\mid x_{i};\hat{\theta}_{1(0)})f_{2}(y_{2i}\mid x_{i},y_{1i};\hat{\theta}_{2(0)})}{\int f_{1}(y_{1i}\mid x_{i};\hat{\theta}_{1(0)})f_{2}(y_{2i}\mid x_{i},y_{1i};\hat{\theta}_{2(0)})dy_{1i}}.\label{6-10}
\end{equation}
Note that the marginal distribution of $x$, $f_x( x; \theta_0)$, is not used in (\ref{6-10}). 
Except for some special cases such as when both $f_{1}$ and $f_{2}$ are
normal distributions, the conditional distribution in (\ref{6-10})
is not in a known form. Thus, some computational tools such as Metropolis-Hasting (\citealt{hastings1970monte}) or  
 SIR (Sampling
Importance Resampling, \citealt{smith92}) are needed to generate samples from
(\ref{6-10}) for $i\in A_{01}$. For example, the SIR consists of the following
steps: 
\begin{enumerate}
\item Generate $B$ (say $B=100$) Monte Carlo samples, denoted by $y_{1i}^{*(1)}, \cdots, y_{1i}^{*(B)}$,  from $f_{1}(y_{1i}\mid x_{i};\hat{\theta}_{1(0)})$. 
\item Among the $B$ samples obtained from  Step 1, select one sample with the selection probability proportional to  $f_{2}(y_{2i}\mid x_{i},y_{1i}^{*(k)};\hat{\theta}_{2(0)})$, where $y_{1i}^{*(k)}$ is the $k$-th sample from Step 1 ($k=1, \cdots, B$). 
\item Repeat Step 1 and Step 2 independently $M$ times to obtain $M$ imputed
values. 
\end{enumerate}
Once we obtain $M$ imputed values of $y_{1i}$, we can use 
\[
{h}(y_{1i}\mid x_{i},y_{2i})\propto f_{1}(y_{1i}\mid x_{i};\hat{\theta}_{1(0)})f_{2}(y_{2i}\mid x_{i},y_{1i};\hat{\theta}_{2(0)})
\]
as  the proposal density in (\ref{6-8}). Since $\sum_{j=1}^{M}w_{ij}^{*}=1$,
we do not need to compute the normalizing constant in (\ref{6-10}). For  $i \in A_{10}$, $M$ imputed values of $y_{2i}$ are generated from $f_2 ( y_{2i} \mid x_i, y_{1i}; \hat{\theta}_{2(0)} )$. For $i \in A_{(00)}$, $M$ imputed values of $y_{1i}$ are generated from $f_1( y_{1i} \mid x_i ; \hat{\theta}_{1(0)} )$ and then $M$ imputed values of $y_{2i}$ are generated from $f_2 ( y_{2i} \mid x_i, y_{1i}^* ; \hat{\theta}_{2(0)} )$. 

In the W-step, the fractional weights are computed by 
\[
w_{ij(t)}^{*}\propto\frac{f_{1}(y_{1i}^{*(j)}\mid x_{i};\hat{\theta}_{1(t)})f_{2}(y_{2i}^{*(j)}\mid x_{i},y_{1i};\hat{\theta}_{2(t)})}{h(y_{mis,i}^{*(j)}\mid x_{i},y_{obs,i})}
\]
with $\sum_{j=1}^{M}w_{ij(t)}^{*}=1$, where $y_{1i}^{*(j)}=y_{1i}$
if $y_{1i}$ is observed and $y_{2i}^{*(j)}=y_{2i}$ if $y_{2i}$
is observed.

%The M-step and C-step remain the same. 

\end{example}

The above example covers a broad range of applications in the
missing data literature, such as missing covariate problems, measurement
error models, generalized linear mixed models, and so on. \citet*{Yang2014SemiparametricInference} considered regression
analyses with missing covariates in survey data using FI, where in the current notation,
$f(y_{2}\mid x,y_{1})$ is a regression model with $y_{2}$ and $x$
 fully observed and $y_{1}$  subject to missingness. 
%\citet*{kimberg2015} considered a statistical matching problem where $y_1$ and $y_2$ are never jointly observed and developed a novel application of FI. 
 In generalized linear mixed models, $f(y_{2}\mid x,y_{1})$
is a generalized linear mixed model where $y_{1}$ is the latent random
effect. See \citet*{Yang2013parametric} for using FI to estimate
parameters in the generalized linear mixed models.

For variance estimation, note that the imputed estimator $\hat{\eta}_{FI}$ obtained from the imputed estimating equation (\ref{1-5}) depends on $\hat{\theta}$ obtained from (\ref{eq:approximation1}). To reflect this dependence, we can write $\hat{\eta}_{FI} = \hat{\eta}_{FI} ( \hat{\theta})$. To account for the sampling variability of $\hat{\theta}$ in the imputed estimator $\hat{\eta}_{FI}$,  
 either the linearization method or replication methods can be used. In the linearization method,  the imputation model is needed in order to compute partial derivatives of the score functions. To avoid disclosing the imputation model, 
 replication methods are often preferred (\citealt{rao1992jackknife}).  To
implement the replication variance estimation in FI, we
first obtain the $k$-th replicate pseudo MLE $\hat{\theta}^{[k]}$
of $\hat{\theta}$ by solving 
\begin{equation}
\bar{S}^{*[k]}(\theta)\equiv\sum_{i\in A}w_{i}^{[k]}\sum_{j=1}^{M}w_{ij}^{*}(\theta)S(\theta;y_{ij}^{*})=0,
\label{13}
\end{equation}
where $w_{i}^{[k]}$ is the $k$-th replication weight and $w_{ij}^*( \theta)$ is defined in (\ref{9}). 
To obtain $\hat{\theta}^{[k]}$ from (\ref{13}),
either EM algorithm or the one-step Newton method can be used.
EM algorithm can be implemented similarly as before. For the one-step Newton method,  we have 
  $$ \hat{\theta}^{[k]} = \hat{\theta}  -  \left\{ \frac{ \partial }{ \partial \theta^T}  \bar{S}^{*[k]}( \hat{\theta} )
   \right\}^{-1}  
  \sum_{i\in A}w_{i}^{[k]}\sum_{j=1}^{M}w_{ij}^{*}(\hat{\theta})S(\hat{\theta};y_{ij}^{*}),
    $$
    where 
    \begin{equation}
    \begin{aligned}
\nonumber   & \frac{ \partial }{ \partial \theta^T}  \bar{S}^{*[k]}( {\theta} ) = 
    \sum_{i\in A}w_{i}^{[k]}\sum_{j=1}^{M}w_{ij}^{*}(\theta)\dot{S}(\theta;y_{ij}^{*})   
 +     \sum_{i\in A}w_{i}^{[k]}\sum_{j=1}^{M}w_{ij}^{*}(\theta) \cdot \\  & \qquad\qquad \left\{ S({\theta};y_{ij}^{*})-   \sum_{j=1}^{M}w_{ij}^{*}(\theta) S({\theta};y_{ij}^{*})  \right\}^{\otimes 2}  
  \end{aligned}
  \end{equation}    
  with $\dot{S}(\theta;y) = \partial S( \theta ; y) / \partial \theta^T$ and $B^{\otimes 2} = B B^T$. 
    Once $\hat{\theta}^{[k]}$ is obtained,  we obtain the $k$-th replicate $\hat{\eta}^{[k]}$
of $\hat{\eta}$ by solving 
\[
\sum_{i\in A}w_{i}^{[k]}\sum_{j=1}^{M}w_{ij}^{*[k]}U(\eta;y_{ij}^{*})=0
\]
for $\eta$, where $w_{ij}^{*[k]}= w_{ij}^* ( \hat{\theta}^{[k]})$. 
%If $M$ is not sufficiently large, calibration fractional imputation method, discussed in Section 6.2, can be used.  

\section{NONPARAMETRIC FRACTIONAL IMPUTATION}

\subsection{Fractional Hot Deck Imputation}

\textit{Hot deck imputation} uses observed responses from the sample
as imputed values. The unit with a missing value is called the\textit{
recipient} and the unit providing the value for the imputation is
called the \textit{donor}. \citet*{durrant2009imputation}, \citet*{haziza2009imputation}
and \citet*{andridge2010review} provided comprehensive overviews of
hot deck imputation in survey sampling. The attractive
features of hot deck imputation include the following.
First, unlike model-based imputation methods that generate artificial imputed values,
in hot deck imputation, only plausible values can be imputed, and
therefore distributional properties of the data are preserved. For
example, imputed values for categorical variables will also be categorical,
as observed from the respondents.
Second, compared to fully parametric methods, hot deck imputation makes
less or no distributional assumptions and therefore  is more robust. For these reasons, hot deck imputation
is a widely used imputation method, especially in household surveys.

\textit{Fractional hot deck imputation} (FHDI) combines the ideas of FI and
hot deck imputation. It is  efficient (due
to FI), and it inherits the aforementioned good properties of hot
deck imputation. \citet*{kim2004fractional}, \citet*{fuller2005hot},  
and \citet*{Kim2014Fractionalhotdeck} considered FHDI for  univariate missing data. 
 We now describe a multivariate
FHDI procedure to deal with missing data with an arbitrary missing pattern
(\citealt{im2015}). 

We first consider categorical data. Let $\mathbf{z}=(z_{1},\ldots,z_{K})$
be the vector of study variables that take categorical values. Let
$\mathbf{z}_{i}=(z_{i1},\ldots,z_{iK})$ be the $i$-th realization
of $\mathbf{z}$. Let $\delta_{ij}$ be the response indicator variable
for $z_{ij}$. That is, $\delta_{ij}=1$ if $z_{ij}$ is observed and $\delta_{ij}=0$ otherwise. 
Assume that the response mechanism is MAR. Based on
$\delta_{i}=(\delta_{i1},\dots,\delta_{iK})$,
the original observation $\mathbf{z}_{i}$ can be decomposed into
$(z_{obs,i},z_{mis,i})$, which are the missing and observed part
of $\mathbf{z}_{i}$, respectively. Let $D_{i}=\{z_{mis,i}^{*(1)},\ldots,z_{mis,i}^{*(M_{i})}\}$
be the set of all possible values of $z_{mis,i}$, that is, $(z_{obs,i},z_{mis,i}^{*(j)})$
is one of the actually observed value in the respondents, for $j=1,\ldots,M_{i}$,
with $M_{i}>0$. If all of $M_{i}$ possible values are taken as the
imputed values for $z_{mis,i}$, the fractional weight assigned to
the $j$-th imputed value $z_{mis,i}^{*(j)}$ is 
\begin{equation}
w_{ij}^{*}=\frac{\pi(z_{obs,i},z_{mis,i}^{*(j)})}{\sum_{k \in D_i}\pi(z_{obs,i},z_{mis,i}^{*(k)})},\label{eq:fhdi weights}
\end{equation}
where $\pi(\mathbf{z})$ is the joint probability of $\mathbf{z}$. If the joint probability
is nonparametrically modeled, it is computed by 
\begin{equation}
\pi(\mathbf{z})=\frac{\sum_{i\in A}w_{i}\sum_{j\in D_{i}}w_{ij}^{*}I\{(z_{obs,i},z_{mis,i}^{*(j)})=\mathbf{z}\}}{\sum_{i\in A}w_{i}},\label{eq:nonparametric}
\end{equation}
where  $z_{mis,i}^{*(j)}\equiv z_{mis,i}$
and $w_{ij}^{*}=M_{i}^{-1},$ for $j=1,\ldots,M_{i}^{-1}$, if $\mathbf{z}_{i}$ is completely observed. To compute
(\ref{eq:fhdi weights}) and (\ref{eq:nonparametric}), EM algorithm by weighting (\citealt{Ibrahim90})
can be used, with the initial values of fractional weights being $w_{ij(0)}^{*}=M_{i}^{-1}$. Equations 
(\ref{eq:fhdi weights}) and (\ref{eq:nonparametric}) correspond
to the E-step and M-step of the EM algorithm, respectively. The M-step
(\ref{eq:nonparametric}) can be changed if there is a parametric
model for the joint probability $\pi(\mathbf{z})$. For example, if the joint probability
can be modeled by a multinomial distribution with parameter $\alpha$, say $\pi(\mathbf{z};\alpha)$,
then the M-step replaces (\ref{eq:nonparametric}) with solving the imputed
score equation of $\alpha$ to update the estimate of $\alpha$.

For continuous data $\mathbf{y}=(y_{1},\ldots,y_{K})$, we consider
a discrete approximation. Discretize each continuous variable by dividing
its range into a small finite number of segments (for example, quantiles).
Let $z_{ik}$ denote the discrete version of $y_{ik}.$ Note that
$z_{ik}$ is observed only if $y_{ik}$ is observed. Let the support
of $\mathbf{z}$, denoted by $\{\mathbf{z}_{1},\ldots,\mathbf{z}_{G}\}$, which
is the same as the sample support of $\mathbf{z}$ from the full respondents, specify donor cells. The joint probability of $\mathbf{z}$,
denoted by $\pi(\mathbf{z}_{g})$, for $g=1,\ldots,G$, can be obtained
by the EM algorithm for categorical missing data as described above. 

As in the categorical missing data problem, let $D_{i}=\{z_{mis,i}^{*(1)},\ldots,z_{mis,i}^{*(M_{i})}\}$
be the set of all possible values of $z_{mis,i}$. 
Using a finite mixture model, a nonparametric approximation of $f(y_{mis,i}\mid y_{obs,i})$
is 
\begin{equation}
f(y_{mis,i}\mid y_{obs,i})\approx\sum_{j=1}^{M_{i}}P(\mathbf{z}= \mathbf{z}_{i}^{*(j)}\mid y_{obs,i})f(y_{mis,i}\mid \mathbf{z}_{i}^{*(j)}).
\label{17}
\end{equation}
Each $\mathbf{z}_{i}^{*(j)}=(z_{obs,i},z_{mis,i}^{*(j)})$ defines
an imputation cell. The approximation in  (\ref{17}) is based on the assumption that 
\begin{equation}
P( y_{mis} \mid  y_{obs},  \mathbf{z} )  \cong  P( y_{mis} \mid   \mathbf{z} ), 
\label{18}
\end{equation}
which requires (approximate) conditional independence between $y_{mis}$ and $y_{obs}$ given $z$.  Thus, we assume that the covariance structure between items  are  captured by the discrete approximation and the within cell errors can be safely assumed to be independent.  Once the imputation cells are formed to satisfy (\ref{18}), we select $m_{g}$
imputed values for $y_{mis,i}$, denoted by $\mathbf{y}_{i}^{*(j)}=(y_{obs,i},y_{mis,i}^{*(j)})$,
for $j=1,\ldots,m_{g}$, randomly from the full respondents
in the same cell, with the selection probability proportional to the
sampling weights. The final fractional weights assigned to
$\mathbf{y}_{i}^{*(j)}$ is $w_{ij}^{*}=\hat{P}(z_{mis,i}^{*(j)}\mid y_{obs,i})m_{g}^{-1}$.

This FHDI procedure resembles a two-phase stratified sampling (\citealt{rao1973double}, \citealt{kim06b}),
where forming the imputation cells corresponds to stratification (phase one) and conducting   hot deck imputation corresponds to stratified sampling (phase two).  For more details, see \citet*{im2015}. 

 If we select all possible donors in the same cell, the resulting FI estimator is fully efficient in the sense that it does not introduce additional randomness due to hot deck imputation. Such fractional hot deck imputation is called fully efficient fractional imputation (FEFI). The FEFI option is currently available at Proc Surveyimpute in SAS (\citealt{SAS2015}). 

\subsection{Nonparametric Fractional Imputation Using Kernels}

In real-data applications, nonparametric  methods are preferred if less is known about the true underlying data model.
Hot deck imputation makes less or no distributional assumptions and therefore
is more robust than fully parametric methods. In what follows, we discuss 
an alternative way of calculating the fractional weights that
links the FI estimator to some well-known nonparametric estimators, such as 
Nadaraya-Watson kernel regression estimator (\citealt{nadaraya1964estimating}). 

For simplicity, suppose we have bivariate data $(x_{i},y_{i})$ where
$x_{i}$ is completely observed and $y_{i}$ is subject to missing.
Assume the missing data mechanism is MAR. Let $\delta_{i}$ be the
response indicator that takes the value one if $y_i$ is observed and takes zero otherwise.  We are interested in estimating $\eta$, which is defined through $ E\{ U( \eta; X, Y) \}=0$. Let $A_R =\{ i \in A; \delta_i = 1 \}$ be the index set of  respondents. 
 To calculate the conditional estimating equation (\ref{eq:condEE}) nonparametrically, we use the following fractional imputation:  for each unit $i$ with $\delta_i=0$, $r = \left| A_R \right|$ imputed values of $y_i$
are taken from $A_R$, denoted by $y_i^{*(1)},  \cdots, y_i^{*(r)}$, and compute  the Kernel-based 
fractional weights 
$w_{ij}^{*}=  K_h (x_{i}-x_i^{*(j)})/\sum_{k\in A_R} K_h( x_{i}-x_i^{*(k)} ) $, where $K_h( \cdot) $ is the kernel function with bandwidth $h$ and $x_{i}^{*(j)}$ is the covariate
associated with $y_{i}^{*(j)}$. 
 The resulting FI estimating equation can be written as
\begin{equation}
\sum_{i \in A} w_i \left\{ \delta_i U( \eta; x_i, y_i) + (1-\delta_i) \sum_{j\in A_R}  w_{ij}^* U( \eta; x_i, y_i^{*(j)})  \right\} =0,
\label{19b}
\end{equation}
 where the nonparametric 
 fractional
weights measure the degrees of similarity based on the distance between
$x_{i}$ and $x_{i}^{*(j)}$.
%where $x_{i}^{*(j)}$ is the covariate
%associated with $y_{i}^{*(j)}$.  
The FI estimator uses $\hat{U}( \eta; x_{i})\equiv\sum_{j\in A_R}w_{ij}^{*} U( \eta; x_i,  y_i^{*(j)} ) $
to approximate  $E\{ U( \eta; x_i, y_{i}) \mid x_i \}$ nonparametrically. For fixed $\eta$,  
$
\hat{U}( \eta; x_{i})$ is often
called the \textit{Nadaraya-Watson kernel regression estimator} of $E\{ U( \eta; x_i, y_{i}) \mid x_i\}$
in the nonparametric estimation framework. Note that this FI estimator does not
rely on any parametric model assumptions and so is nonparametric; however it is not assumption free because it
makes an implicit assumption of the continuity of $E\{ U(\eta;x,y)\mid x_i\}$ through the choice of kernels to define
the ``similarity'' (\citealt{nadaraya1964estimating}). Notably, while the convergence of $\hat{U}( \eta; x_{i})$ to $E\{ U( \eta; x_i, y_{i}) \mid x_i\}$ does not achieve the order  of $O_p(1/\sqrt{n})$, the solution $\hat{\eta}_{FI}$ to (\ref{19b}) satisfies $\hat{\eta}_{FI} - \eta = O_p(1/\sqrt{n})$ under some regularity  conditions, which was proved by \citet*{wang2009empirical} in the IID setup. 

Such kernel-based nonparametric fractional imputation can be directly applicable to complex survey sampling scenarios.
%Replication-based variance estimation may work well with FI, which is under development. 
More developments are expected by coupling FI with other nonparametric methods  such as those using   the nearest neighbor imputation  method (\citealt{chen2001jackknife};
\citealt{kitamura2009variance}; \citealt{kim2011variance})  or predictive mean matching 
(\citealt{vink2014predictive}).
%(\citealt{rubin1976inference}, p 168). 

\section{SYNTHETIC DATA IMPUTATION}
\label{proj3}

Synthetic imputation is a technique of creating imputed values for the unobserved items by incorporating information from other surveys.
 For example, suppose that there are two independent surveys, called Survey 1 and Survey 2, and we observe $x_i$ from Survey 1 and observe $(x_i,y_i)$ from Survey 2. In this case, we may want to create synthetic values of $y_i$ in Survey 1 by first fitting a  model relating $y$ to $x$ to the data from Survey 2 and then predicting $y$ associated with $x$ observed in Survey 1. Synthetic imputation  is particularly useful when Survey $1$ is a large scale survey and item $y$ is very expensive to measure.  \citet*{schenker07} reported several applications of synthetic imputation, using a model-based method to estimate  parameters associated with variables not observed in Survey 1 but observed in a much smaller Survey 2. In one application,  both self-reported health measurements $x_i$ and clinical measurements from physical examinations $y_i$ for a  small sample $A_2$ of individuals were observed. In the much larger Survey 1, only self-reported measurements, $x_i$ were observed. Only the imputed or synthetic data from Survey 1 and associated survey weights were released  to the public.

The setup of two independent samples with common items is often called non-nested two-phase sampling.  Two-phase sampling   can be treated as a missing data problem, where the missingness is planned and the response probability is known.

%Two-phase sampling theory can be extended to the situation where two different sample selections are made with different measurements in each sample. In the measurement error problem, a validation sample can be regarded as a second-phase sample in two-phase sampling. In missing data, the set of respondents can be regarded as a second-phase sample.
%Two-phase sampling

\subsection{Fractional Imputation for  Two-phase Sampling}

In  two-phase sampling, suppose
we observe ${x}_i$ in the first-phase sample and observe $({x}_i, y_i)$ in the second-phase sample, where the second-phase sample
 is not necessarily nested within the first-phase sample. Let $A_1$ and $w_{i1}$ be the set of indices and the set of sampling weights for the first-phase sample, respectively. Let  $A_2$ and $w_{i2}$ be the corresponding sets  for the second-phase sample. Assume a  ``working'' model $m({x}_i; \beta)$ for  $E(y \mid {x}_i)$.
%and for the second-phase sample, respectively.
%Also, let  be the sampling weight of unit $i$ for the first-phase sample and for the second-phase sample, respectively.
For estimation of the population total of $y$, the two-phase regression estimator can be written as
\begin{equation}
\hat{Y}_{tp} = \sum_{i \in A_1} w_{i1} m ( {x}_i; \hat{\beta} )  +
\sum_{i \in A_2 } w_{i2} \{  y_i -m ( {x}_i; \hat{\beta} )  \},
\label{2-4-1}
\end{equation}
 where the subscript "tp" stands for "two-phase", and $\hat{\beta}$ is estimated from the second-phase sample. The two-phase regression estimator is efficient if the working model is well-specified.  The first term of (\ref{2-4-1}) is called the projection estimator. Note that if the second term of (\ref{2-4-1}) is equal to zero, the two-phase regression estimator is equivalent to the projection estimator.
  Some asymptotic properties of the two-phase estimator and variance estimation methods have been discussed in \citet*{kim06b}, and  \citet*{kimyu11b}. \citet*{kimrao12} discussed asymptotic properties of the projection estimator under  non-nested two-phase sampling.

%Kim, J.K. and Yu, C.L.  (2011). ``Replication variance estimation under  two-phase  sampling,''  \emph{Survey Methodology}, {\bf 37}, 67--74.

In a large scale survey, it is a common practice to produce estimates for  domains.  Creating an imputed data set for the first-phase sample, often called mass imputation, is one method for incorporating the second-phase information into  the first-phase sample.  \citet*{breidt96}
discussed the possibility of using  imputation  to get  improved estimates for domains. \citet*{fuller03} investigated  mass imputation in the context of two-phase sampling.
%\citet*{kimrao12} also considered the projection domain estimation.

The FI procedure can be used to obtain the two-phase regression estimator in (\ref{2-4-1}) and, at the same time, improve domain estimation.
Note that the two-phase regression estimator (\ref{2-4-1})  can be written as
\begin{equation}
 \hat{Y}_{FEFI} = \sum_{i \in A_1} \sum_{j \in A_2} w_{i1} w_{ij}^* y_{i}^{*(j)},
 \label{2-4-2}
 \end{equation}
where $y_{i}^{*(j)} =  \hat{y}_i + \hat{e}_j $, $\hat{y}_i=  m ( {x}_i; \hat{\beta} )$, $\hat{e}_j = y_j - \hat{y}_j $,
$ w_{ij}^* = w_{j2}/(\sum_{k \in A_2} w_{k2} ) ,$
 and  we assume $\sum_{i \in A_{1} } w_{i1} = \sum_{i \in A_2} w_{i2} $. The expression (\ref{2-4-2}) implies that we impute all the elements in the first-phase sample, including the elements that also belong to the second-phase sample. The estimator (\ref{2-4-2}) is computed using an augmented data set of $n_1 \times n_2$ records,  where $n_1$ and $n_2$  are the sizes of $A_1$ and $A_2$, respectively, and the  $\left(i,j \right)$-th record has an (imputed) observation $y_{i}^{*(j)} = \hat{y}_i+ \hat{e}_j$ with weight $w_{i1} w_{ij}^* $. That is, for each unit $i \in A_1$, we impute $n_2$ values of  $y_{i}^{*(j)}$ with fractional weight $w_{ij}^*$. The  method in (\ref{2-4-2})  imputes all the elements in $A_2$ and is called  fully efficient fractional imputation (FEFI) method, according to \citet*{fuller2005hot}. The FEFI estimator is algebraically equivalent to the two-phase regression estimator of the population total of $y$,  and  can  also provide consistent  estimates for other parameters such as population quantiles.

If it is desirable to limit the number of imputations to a small value $m$ ($m<n_2$), FI using the regression weighting method in \citet*{fuller2005hot} can be adopted. We first select $m$ values of $y_{i}^{*(j)}$, denoted by $y_{i}^{**(1)}, \cdots, y_{i}^{**(m)}$,  among the set of $n_2$ imputed values $\{ y_{i}^{*(j)}; j \in A_2  \}$ using an efficient sampling method. The fractional weights $\tilde{w}_{ij}^*$ assigned to the selected $y_{i}^{**(j)}$ are determined so that
\begin{equation}
\sum_{j=1}^m \tilde{w}_{ij}^{*} \left( 1 , y_{i}^{**(j)} \right) = \sum_{j \in A_2} w_{ij}^* \left( 1, y_{i}^{*(j)} \right)
\label{2-4-3}
\end{equation}
holds
for each $i \in A_1$.
The fractional weight satisfying (\ref{2-4-3}) can be computed using the regression weighting method or the empirical likelihood method, see section 6.1 for details. The resulting FI data $ y_{i}^{**(j)}$ with weights $w_{i1} \tilde{w}_{ij}^{*} $ are constructed with $n_1 \times m$ records, which integrate available information from two phases.  Replication variance estimation with FI, similar to \citet*{fuller2005hot}, can be developed.  See Section 8.7 of \citet*{kim2013statistical}. 

\subsection{Fractional Imputation for Statistical Matching}

Statistical matching is used to integrate  two or more data sets when information available for matching records for individual participants across data sets is incomplete.   Statistical matching can be viewed as a missing data problem where a researcher wants to perform a joint analysis of variables not jointly observed. Statistical matching techniques can be used to construct fully augmented data files to enable statistically valid data analysis.

\begin{table}[!hbtp]
  \caption{\label{table1} A Simple Data Structure for Matching}
  \begin{center}
  \begin{tabular}{rccc}
      \hline%
       & $X$ & $Y_1$ & $Y_2$ \\
       \cline{2-4}
  Sample A & o & o &  \\
  Sample B & o &   &  o  \\
  \hline
\end{tabular}
  \end{center}
  \label{table9.1}
\end{table}

 To simplify the setup, suppose that there are two surveys, Survey A and Survey B, each containing a random sample with partial information about the population. Suppose that we observe $x$ and $y_1$ from the Survey A sample and observe $x$ and $y_2$ from the Survey B sample.
Table \ref{table9.1} illustrates a simple data structure for matching.

Without loss of generalizability, consider imputing $y_1$ in Survey B, since imputing $y_2$ in Survey A is symmetric. 
Under this setup, we can use FI to  generate $y_1$ from the conditional distribution of $y_1$ given the observations. That is, we  generate $y_1$ from
\begin{equation}
 f\left( y_1 \mid x, y_2 \right) \propto f \left( y_2 \mid x, y_1 \right) f \left( y_1 \mid x \right) .
 \label{9-2}
 \end{equation}
 Of note, assumptions are needed to identify the parameters in the joint model. For example, 
  \citet*{kimberg2015} used an instrumental variable assumption to identify the  model.  
To generate $y_1$ from (\ref{9-2}), the  EM algorithm by FI can be used.
% can be used as follows:
%\begin{description}
%\item [\textit{I-step}] For each $i \in B$,
%generate $m$ imputed values of $y_{1i}$, denoted by $y_{1i}^{*(1)}, \cdots, y_{1i}^{*(M)}$,
%from $ \hat{f}_A \left( y_1 \mid x_i \right)$, where $\hat{f}_A \left( y_1 \mid x \right)$ denotes the estimated density for the conditional distribution of $y_1$ given $x$ from Survey A.
%\item [\textit{W-step}] Let $\hat{\theta}_{(t)}$ be the current parameter value of $\theta$ in
%$f\left( y_2 \mid x_1, y_1 \right)$. For the $j$-th imputed value $y_{1i}^{*(j)}$, assign a fractional weight
%$$ w_{ij(t)}^* \propto f\left( y_{2i} \mid x_{i}, y_{1i}^{*(j)}; \hat{\theta}_{(t)} \right)$$
%with the constraint of $ \sum_{j=1}^M w_{ij(t)}^* = 1$.
%\item [\textit{M-step}] Solve the fractionally imputed score equation for $\theta$
%$$ \sum_{i \in B} w_{ib} \sum_{j=1}^M  w_{ij(t)}^* S(\theta; x_{i},  y_{1i}^{*(j)} , y_{2i}) = 0 $$
%to obtain $\hat{\theta}_{(t+1)}$, where $S(\theta; x, y_1, y_2) = \partial \log f ( y_2 \mid x, y_1; \theta)/\partial \theta$.
%\item [\textit{Iteration}]  Go to the W-step and continue until convergence.
%\end{description}
%The above EM algorithm constructs the fractionally imputed values of $y_{1i}$ and the %associated fractional weights in Survey B.  The imputed data can be treated as complete data, %which enables joint analyses of $y_1$ and $y_2$. The user can  apply standard parameter %estimation to the fractionally imputed data. 
For more details, see \citet*{kimberg2015}. 
% Variance estimation of the FI estimator needs to be developed.

\section{FRACTIONAL IMPUTATION VARIANTS}

\subsection{The Choice of $M$ and Calibration Fractional Imputation}

The choice of the imputation size $M$ is a matter of tradeoff between statistical efficiency
and computation efficiency: small $M$ may lead to large variability
in Monte Carlo approximation; whereas large $M$ may increase computational
cost. The magnitude of the imputation error is usually $O(1/\sqrt{M})$, which
can be reduced for large $M$. Thus, if computational power allows,
the larger $M$, the better.

In survey practices, a large imputation size
may not be desirable. Thus, instead of releasing to public large number
of imputed values for each missing item, a subset of initial imputation
values can be selected to reduce the imputation size.
In this case, the FI procedure can be developed in  three stages. The
first stage, called \textit{Fully Efficient Fractional Imputation} (FEFI),
computes the pseudo MLE of parameters in the superpopulation
model with sufficiently large imputation size $M$, say $M=1,000$. 
The second stage
is the \textit{Sampling} Stage, which selects small $m$ (say, $m=10$) imputed values from the set
of $M$ imputed values. The third stage is \textit{Calibration
Weighting}, which involves constructing the final fractional weights
for the $m$ final imputed values to satisfy some calibration constraints. This procedure can be called \textit{Calibration
FI}.

The FEFI step is the same as in the previous section. In what follows,
we describe the last two stages in details. In the Sampling Stage, 
 a subset of imputed values are selected to reduce the imputation size.
For each $i$, we have $M$ imputed values $y_{ij}^{*}=(y_{obs,i},y_{mis,i}^{*(j)})$
with their fractional weights $w_{ij}^{*}$. We treat $\mathbf{y}_{i}^{*}=\{y_{ij}^{*},j=1,\ldots,M\}$
as a weighted finite population with weight $w_{ij}^{*}$ and use
an unequal probability sampling method such as probability-proportion-to-size (PPS) sampling to select a sample of size $m$, say $m=10$,
from $\mathbf{y}_{i}^{*}$ using $w_{ij}^{*}$ as the selection probability.
Let $\tilde{y}_{i1}^{*},\ldots,\tilde{y}_{im}^{*}$ be the $m$ elements
sampled from $\mathbf{y}_{i}^{*}$. 

The initial  fractional weights for the sampled $m$ imputed values are given
by $\tilde{w}_{ij0}^{*}=m^{-1}.$ This set of fractional weights may
not necessarily satisfy the imputed score equation
\begin{equation}
\sum_{i\in A}w_{i}\sum_{j=1}^{m}\tilde{w}_{ij}^{*}S(\hat{\theta};\tilde{y}_{ij}^{*})=0,\label{eq:CalibrationFI}
\end{equation}
where $\hat{\theta}$ is the pseudo MLE of $\theta$ computed at the
FEFI stage.  It is desirable for  the solution to the imputed score equation with small $m$ to be equal to the pseudo MLE of $\theta$, which specifies the calibration constraints.  At the Calibration Weighting
stage, the initial set of weights are modified to satisfy the constraint
(\ref{eq:CalibrationFI}). Finding the calibrated fractional weights
can be achieved by the regression weighting technique, by which the
fractional weights that satisfy (\ref{eq:CalibrationFI}) and $\sum_{j=1}^{m}\tilde{w}_{ij}^{*}=1$. The regression fractional weights
are constructed by 
\begin{equation}
\tilde{w}_{ij}^{*}=\tilde{w}_{ij0}^{*}+\tilde{w}_{ij0}^{*}\Delta(S_{ij}^{*}-\bar{S}_{i}^{*}),\label{eq:calibration weights}
\end{equation}
where $S_{ij}^{*}=S(\hat{\theta};y_{ij}^{*})$, $\bar{S}_{i}^{*}=\sum_{j=1}^{m}\tilde{w}_{ij0}^{*}S_{ij}^{*},$ and
\[
\Delta=-\{\sum_{i\in A}w_{i}\sum_{j=1}^{m}\tilde{w}_{ij0}^{*}S_{ij}^{*}\}^{T}\{\sum_{i\in A}w_{i}\sum_{j=1}^{m}\tilde{w}_{ij0}^{*}(S_{ij}^{*}-\bar{S}_{i}^{*})^{\otimes 2}\}^{-1}.
\]
Note that some of the fractional weights computed by (\ref{eq:calibration weights})
can take negative values. To avoid negative weights,  alternative
algorithms other than  regression weighting should be used. For
example, the fractional weights of the form
\[
\tilde{w}_{ij}^{*}=\frac{\tilde{w}_{ij0}^{*}\exp(\Delta S_{ij}^{*})}{\sum_{k=1}^{m}\tilde{w}_{ik0}^{*}\exp(\Delta S_{ik}^{*})}
\]
are approximately equal to the regression fractional weights in (\ref{eq:calibration weights})
and are always positive.

\subsection{The Choice of the Proposal Distribution\label{sub:Choice of h}}

PFI is based on  sampling from an importance sampling  density $h$
called the \textit{proposal distribution}. The choice of the proposal
distribution is somewhat arbitrary. However, with finite
samples and imputations, a well-specified proposal distribution may
improve the performance of the imputation estimator. There are a number
of ways to specify the proposal distribution and to assess the goodness
of specification.

For a planned parameter, e.g., $\eta$, the population mean of $y$,
\citet*{kim11} showed the optimal $h^{*}$ that makes
Monte Carlo approximation variance of $\bar{y}_{i}^{*}\equiv\sum_{j=1}^{M}w_{ij}^{*}y_{ij}^{*}$
as small as possible, is given by 
\[
h^{*}(y_{mis,i}|y_{obs,i})=f(y_{mis,i}|y_{obs,i},\hat{\theta})\times\frac{|y_{i}-\mathbb{E}\{y_{i}|y_{obs,i},\hat{\theta}\}|}{\mathbb{E}\{|y_{i}-\mathbb{E}\{y_{i}|y_{obs,i},\hat{\theta}\}|y_{obs,i},\hat{\theta}\}},
\]where $\hat{\theta}$ is the MLE of $\theta$.
For general-purpose estimation, $\eta$ is often unknown at the time of
imputation according to \citet*{fay1992inferences}, $h(y_{mis,i}|y_{obs,i})=f(y_{mis,i}|y_{obs,i};\hat{\theta})$
is a reasonable choice in terms of statistical efficiency. For importance
sampling, since we do not know $\hat{\theta}$ at the outset of the EM algorithm,
we may want to have a good initial guess $\theta_{0}$ and use $h(y_{mis,i}|x_{i},y_{obs,i})=f(y_{mis,i}|x_{i},y_{obs,i};\theta_{0})$. If we don't have a good initial guess of the true value of
$\theta$, we can use a prior distribution $\pi(\theta)$ to get $h(y_{mis,i}|y_{obs,i})=\int f(y_{mis,i}|y_{obs,i};\theta)\pi(\theta)d\theta$.

We now discuss  a special choice of the proposal distribution
$h$, based on the realized values of the variables having
missing values, which is akin to hot deck imputation. Without loss of generality, assume that $y_i$ is observed in the first $r$  elements, $y_i$ is missing in the remaining $(n-r)$ elements, and $x_i$ is completely observed in the sample.  
Using the importance
sampling idea, we assign a fractional weight to donor $y_{j}$
($1\leq j\leq r$) for the missing item $y_{i}$ ($r+1\leq i\leq n$)
by choosing $h(y_{j})=f(y_{j}\mid\delta_{j}=1)$. In calculating
the fractional weights, we approximate $f(y_{j}\mid\delta_{j}=1)$
by its empirical distribution $n_{R}^{-1}\sum_{k=1}^{N}\delta_{k}f\left(y_{j}\mid x_{k}\right)$,
where $n_{R}$ is the number of respondents. The EM algorithm takes
the following steps:
\begin{description}
\item [\textit{{I-step}}] For each missing value $y_{i}$, $i=r+1,\ldots,n$, take
all values in $A_{R}=\{y_{1},\ldots,y_{r}\}$ as donors. 
\item [\textit{{W-step}}] With the current estimate of $\theta$, denoted by $\hat{\theta}_{(t)},$compute
the fractional weights by 
\begin{equation}
w_{ij(t)}^{*}\propto\frac{f(y_{j}\mid x_{i};\hat{\theta}_{(t)})}{\sum_{k\in A_{R}}w_{k}f(y_{j}\mid x_{k};\hat{\theta}_{(t)})}\label{fwgt}
\end{equation}

\item [\textit{{M-step}}] Update the parameter $\hat{\theta}_{(t+1)}$ by solving
the following imputed score equation, 
\[
\hat{\theta}_{(t+1)}:\text{solution to }
\sum_{i=1}^rS(\theta;x_i,y_i)+\sum_{i=r+1}^n\sum_{j=1}^rw_{ij}^{*(t)}S(\theta;x_i,y_j)=0.
%\sum_{i\in A}w_{i}\sum_{j=1}^{M}w_{ij}^{*(t)}S(\theta;x_{i},y_{j})=0.
\]

\item [\textit{{Iteration}}] Set $t=t+1$ and go to the W-step. Stop if $\hat{\theta}_{(t+1)}$
meets the convergence criterion. 
\end{description}
The semiparametric fractional imputation (SFI) estimator of $\bar{Y}$ is 
\[
\hat{\bar{Y}}_{SFI}=\frac{1}{n}\left\{ \sum_{i=1}^{r}y_{i}+\sum_{i=r+1}^{n}\sum_{j=1}^{r}w_{ij}^{*}y_{j}\right\} .
\]
\citet*{Kim2014Fractionalhotdeck} showed that the resulting estimator gains robustness. It is less sensitive against
the departure from the assumed conditional regression model. 

\subsection{Doubly Robust Fractional Imputation}

%Doubly robustness estimation is also a popular topic in missing data analysis. 

Suppose we have bivariate data $(x_{i},y_{i})$ where $x_{i}$ is
completely observed and $y_{i}$ is subject to missing and missing data
mechanism is MAR. 
Assume also an outcome regression
(OR) model, given by $E(y_{i}\mid x_{i})=m(x_{i};\beta_{0})$, and
the response propensity (RP) model, given by $P(\delta_{i}=1\mid x_{i},y_{i})=P(\delta_{i}=1\mid x_{i})=\pi(x_{i};\phi_{0})$.
%By MAR, the RP model and $\beta_{0}$
%separate out in the likelihood function, therefore the efficiency with which we can estimate $\beta_{0}$ or any parameter of interest as a function of $\beta_{0}$
%with large samples is the same regardless of whether the treatment initiation model is unknown or parametrized. 
Denote the set of respondents as $A_{R}=\{{i},\delta_{i}=1\}$,
where $\delta_{i}$ is the response indicator of $y_i$. We are interested in
the population total $\eta = \sum_{i=1}^N y_i$. 
Note that not both the OR and RP models are needed to construct consistent estimators of $\eta$. For example, $\hat{\eta}_1=\sum_{i \in A} w_i  m(x_i;\hat{\beta})$, with $\hat{\beta}$ being a consistent 
estimator of $\beta_0$, is consistent to $\eta$ under the OR model and $\hat{\eta}_2= \sum_{i\in A_R} w_i  y_i/\pi(x_i;\hat{\phi})$, with 
 $\hat{\phi}$ being a consistent 
estimator of $\phi_0$, is consistent to $\eta$ under the RP model.

An estimator of $\eta$ is doubly robust if it is consistent if either
the OR model or the RP model is correct, but not necessarily both.
This property guards the estimator from possible model
misspecifications. The DR estimators have been extensively studied
in the literature, including \citet*{robins1994estimation}, \citet*{bang2005doubly}, \citet*{tan2006distributional},
\citet*{kang2007demystifying}, \citet*{cao2009improving}, and \citet*{kim2013doubly}.
We now discuss a fractional imputation estimator that has the double
robustness feature. 

For each missing $y_{i}$, let $y_{ij}^{*}=\hat{y}_i+\hat{e}_{j}$ be the $j$-th imputed value from the donor $j\in A_{R}$, where
$\hat{y}_i=m(x_{i};\hat{\beta})$ with $\hat{\beta}$ fitted under the OR model and $\hat{e}_{j}=y_{j}-m(x_{i};\hat{\beta})$.
If $\sum_{ i \in A_{R}} w_i 1/\pi(x_{j};\hat{\phi})= \sum_{ i \in A} w_i $,
each unit $j\in A_{R}$ represents $1/\pi(x_{j};\hat{\phi})$ copies
of the sample. Then, the fractional weight $w_{ij}^{*}$ associated
with the $j$-th imputed value $y_{ij}^{*}$ is proportional to $\{1/\pi(x_{j};\hat{\phi})-1\}$
over the donor pool $A_{R}$ (minus one because $y_{j}$ itself counts
one), that is, 
\begin{equation}
w_{ij}^{*}= \frac{w_j \{ 1/\pi(x_{j};\phi_{0})-1 \}}{\sum_{k \in A} w_k\delta_{k}\{1/\pi(x_{k};\hat{\phi})-1\}}.\label{eq:dr_weight}
\end{equation}
Under this weight construction, the fractional imputation estimator
is given by
\begin{equation}
\hat{\eta}_{FI}=\sum_{i \in A} w_i \left[  \delta_{i}y_{i}+ (1-\delta_{i})\{\sum_{j=1}^{n}\delta_{j}w_{ij}^{*}y_{ij}^{*}\}\right].\label{eq:DRFI estimator}
\end{equation}
We show that the fractional
imputation estimator $\hat{\eta}_{FI}$ in (\ref{eq:DRFI estimator})
is doubly robust. First notice that $\hat{\eta}_{FI}$ is algebraically
equal to 
\begin{equation}
\hat{\eta}_{FI}= \sum_{i \in A} w_i \left[ m(x_{i};\hat{\beta})+ \frac{\delta_{i}}{\pi(x_{i};\hat{\phi})}\{y_{i}-m(x_{i};\hat{\beta})\}\right].\label{eq:DR estimator}
\end{equation}
Let $\hat{\eta}_{n}=\sum_{i \in A} w_i y_{i}$ be the full sample estimator of 
of  $\eta$, then 
\[
\hat{\eta}_{FI}-\hat{\eta}_{n}=\sum_{i \in A} w_i  \left\{ \frac{\delta_{i}}{\pi(x_{i};\hat{\phi})}-1 \right\}\{y_{i}-m(x_{i};\hat{\beta})\}.
\]
This is an asymptotically unbiased estimator of zero if either the OR model or
the RP model is correct, but not necessarily both. \citet*{kim2013doubly} discussed efficient estimation of $(\beta, \phi)$ in survey sampling.

\section{COMPARISON WITH MULTIPLE IMPUTATION}

\subsection{Statistical Efficiency}

In the presence of missing data with MAR, multiple imputation (MI)
is a popular method. 
%more widely used than FI. 
It is thus
of interest to compare the behavior of these two methods. We start from  a simple setting
with the complete data $z$ being randomly drawn from a population
whose density is $f(z;\theta)$, where $\theta\in\mathbb{R}^{d}$
is an unknown parameter to be estimated. Suppose that $m$ complete
data sets are created by imputing the missing data $z_{mis}$ from the posterior predictive distribution given the observed
data $z_{obs}$
$f(z_{mis}\mid z_{obs})=\int f(z_{mis}\mid z_{obs};\theta)\pi(\theta\mid z_{obs})d\theta$,
where $\pi(\theta\mid z_{obs})$ is the posterior distribution of
$\theta$. The MI estimator of $\theta$, denoted by $\hat{\theta}_{MI}$
is 
\[
\hat{\theta}_{MI}=m^{-1}\sum_{k=1}^{m}\hat{\theta}^{(k)},
\]
where $\hat{\theta}^{(k)}$ is the MLE estimator applied to the $k$-th
imputed data set. Rubin's formula is used for variance estimation
in MI, 
\[
\hat{V}_{MI}(\hat{\theta}_{MI})=W_{m}+(1+m^{-1})B_{m},
\]
where $W_{m}=m^{-1}\sum_{k=1}^{m}\hat{V}^{(k)}$, $B_{m}=(m-1)^{-1}\sum_{k=1}^{m}(\hat{\theta}^{(k)}-\hat{\theta}_{MI})^{2}$,
and $\hat{V}^{(k)}$ is the variance estimator of $\hat{\theta}$
under complete response applied to the $k$-th imputed data set.

Of note, Bayesian MI is a simulation-based method and thus introduce additional
noise. This explains why the asymptotic variance of the MI estimator,
given by \citet*{wang1998large}, 
\begin{equation}
V_{MI}=\mathcal{I}_{obs}^{-1}+m^{-1}\mathcal{I}_{com}^{-1}\mathcal{I}_{mis}\mathcal{I}_{com}^{-1}+m^{-1}J^{T}\mathcal{I}_{obs}^{-1}J,\label{eq:V_MI}
\end{equation}
 is strictly larger than the asymptotic variance
of the FI estimator
\begin{equation}
V_{FI}=\mathcal{I}_{obs}^{-1}+m^{-1}\mathcal{I}_{com}^{-1}\mathcal{I}_{mis}\mathcal{I}_{com}^{-1}, \label{eq:V_FI}
\end{equation}
 where $\mathcal{I}_{com}=E\{S(\theta)^{\otimes2}\}$,
$\mathcal{I}_{obs}=E\{S_{obs}(\theta)^{\otimes2}\}$ , $\mathcal{I}_{mis}=\mathcal{I}_{com}-\mathcal{I}_{obs}$,
$S(\theta)=S(Z;\theta)=\partial\log f(Z;\theta)/\partial\theta$ is
the log likelihood score if the data were completely observed and
$S_{obs}(\theta)=E\{S(\theta)\mid Z_{obs}\}$ is the score function
of the observed data log likelihood , $J=\mathcal{I}_{mis}\mathcal{I}_{com}^{-1}$
is the fraction of missing information matrix (\citealt{rubin1987multiple},
Chapter 4). This difference between (\ref{eq:V_MI}) and (\ref{eq:V_FI})
can be sizable for a small $m$. Furthermore, for a large $m$, although
the MI estimator is efficient, the inference is inefficient
since Rubin's variance estimator of the MI estimator is only
weakly unbiased, that is $\hat{V}_{MI}(\hat{\theta}_{MI})$ converges
in distribution 
instead of coverages in probability to $V_{MI}$. This leads to much broader
confidence intervals and less powerful tests than a consistent variance
estimator would do (\citealt{nielsen2003proper}). 

For MI inference to be valid for general-purpose estimation, imputations
must be proper according to \citet*{rubin1987multiple}. A sufficient condition
is given by \citet*{meng1994multiple}. The so-called congeniality
condition, imposed on both the imputation model and the form of subsequent
complete-sample analyses, is quite restrictive for general-purpose
estimation. Otherwise, as discussed by Fay \citeyearpar{fay1992inferences,fay1996alternative},
\citet*{kott1995paradox}, \citet*{binder1996frequency}, \citet*{robins2000inference},
\citet*{nielsen2003proper}, and \citet{kim2006bias}, the MI variance estimator is not always consistent. \citet{kim2011variance}
pointed out that MI that is congenial for mean estimation is not necessarily
congenial for proportion estimation. \citet*{yang2015mi} showed that
the MI variance estimator can be positively or negatively biased
when the method of moments estimator is used as the complete-sample
estimator. In contrast, FI, as we discussed in section 4, does not require congeniality and
always results in a consistent variance estimator
for general-purpose estimation.

\subsection{Imputation under Informative Sampling}

Under informative sampling, the MAR assumption is subtle. We assume
that the response mechanism is MAR at the population level, now referred
to as population missing at random (PMAR), to be distinguished from the
concept of sample missing at random (SMAR). For simplicity, assume
$y$ is a one-dimensional variable which is subject to missing, $\delta$
is its response indicator, and $I$ is the sample inclusion indicator. PMAR
assumes that $y\perp\delta\mid x$, that is, MAR holds at the population
level, $f(y\mid x)=f(y\mid x,\delta)$. On the other hand, SMAR assumes
$Y\perp\delta\mid(x,I=1)$, that is, MAR holds at the sample level,
$f(y\mid x,I=1)=f(y\mid x,I=1,\delta)$. The two assumptions are not
testable empirically. The plausibility of these assumptions should
be judged by subject matter experts. Often, PMAR is more realistic
because an individual's decision on whether or not to respond to a
survey depends on his or her own characteristics, rather than
the fact of him or her being in the sample or not. 

For noninformative sampling design, we have 
 $P(I=1\mid x,y)=P(I=1\mid x)$, under which PMAR
implies SMAR; however for {informative} sampling design, PMAR
does not necessarily imply SMAR. In such cases, using an imputation model fitted to the sample data for generating imputations
can result in biased estimation. 

FI does not require SMAR to hold besides PMAR. Under PMAR, we have
$f(y\mid x,\delta=0)=f(y\mid x)$. Let $f(y\mid x;\beta)$ be a parametric
model of $f(y\mid x)$. The parameter $\beta$ can be consistently estimated
by solving (\ref{eq:mean-score}), even under informative sampling. Since
FI generates the imputations from $f(y\mid x;\hat{\beta})$, with a consistent estimator $\hat{\beta}$,
the resulting FI estimator is approximately unbiased (\citealt{berg2015}). 
Whereas, MI tends to problematic under  informative sampling.
By using an augmented model, where the imputation model is augmented
to include sampling weights or some function of them,
as $f(y\mid x,w)$, the MI point estimator was claimed to be approximately unbiased
(\citealt{rubin1996multiple,schenker2006multiple}). However, 
 as pointed out by 
\citet*{berg2015}, it is not always true. For example, $Y$ is conditionally independent of $\delta$ given $X$  as presented in Figure 1. However, $Y$ is not conditionally independent of $\delta$ given $X$ and $I$. Augmenting $X$ by including sampling weights does not solve the problem. The existence of the latent variable $U$, which is correlated with $I$ and $\delta$, makes SMAR unachievable. 

\begin{figure}
\begin{center}
\begin{tikzpicture}[->,>=stealth',shorten >=1pt,auto,node distance=3cm,
  thick,main node/.style={circle, draw,font=\sffamily\Large\bfseries}]
  
  \node[main node] (X) {$X$};
  \node[main node] (Y) [left of=X] {$Y$};
  \node[main node](U)[right of = X]{$U$};
  \node[main node](R)[above left of =X]{$\delta$};
 \node[main node](I)[above right of = X]{$I$};

  \path[every node/.style={font=\sffamily\small}]
  (X) edge  node   [left] { }  (Y)	
  (X) edge node   [left]  { } (R) 
  (X) edge node   [left]  { } (I) 
    (Y) edge node  [left]  { } (I)  
  (U) edge node [left]  {  } (I)
  (U) edge node [left]  { } (R) ;
  
\end{tikzpicture}
\end{center} 
\label{figure1} 
\caption{A 
directed acyclic graph
(DAG) for a setup where PMAR holds but SMAR does not hold. Variable $U$ is latent in the sense that it is never observed. } 
\end{figure}
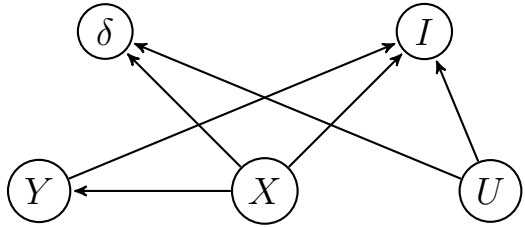

\section{SIMULATION STUDY }

We investigated the performance of FI compared to MI by a limited
simulation study using an artificial finite population generated from
real survey data. The pseudo finite population was generated from
a single month of the U.S. Census Bureau's Monthly Retail Trade Survey
(MRTS). Each month, the MRTS surveys a sample of about $12,000$ retail
businesses with paid employees to collect data on sales and inventories.
The MRTS is an economic indicator survey whose monthly estimates are
inputs to the Gross Domestic Product estimates. The MRTS sample design
is typical of business surveys, employing one-stage stratified sampling
with stratification based on major industry, further substratified
by the estimated annual sales. The sample design requires higher sampling
rates in strata with larger units than in strata with smaller units. 
More details about MRTS can be found in \citet*{mulry2014detecting}.

\begin{figure}
\begin{centering}
\includegraphics[width=5.5in, height=3.8in, scale=0.5]{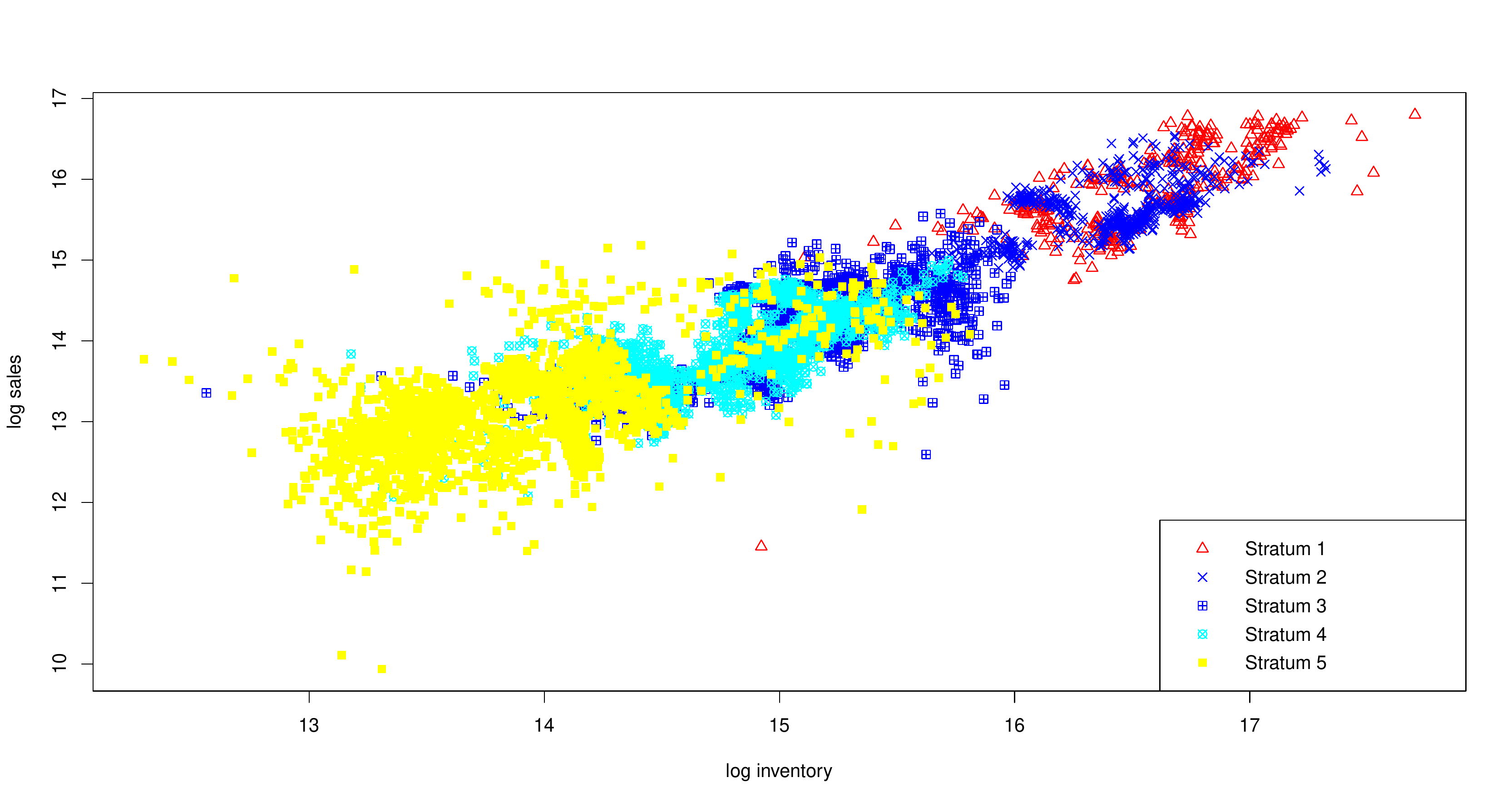} 
\par\end{centering}

\protect\protect\caption{\label{fig:box plot}Scatter plot of log sales and log inventory data by strata }
\end{figure}

\begin{figure}
\begin{centering}
\includegraphics[scale=0.48]{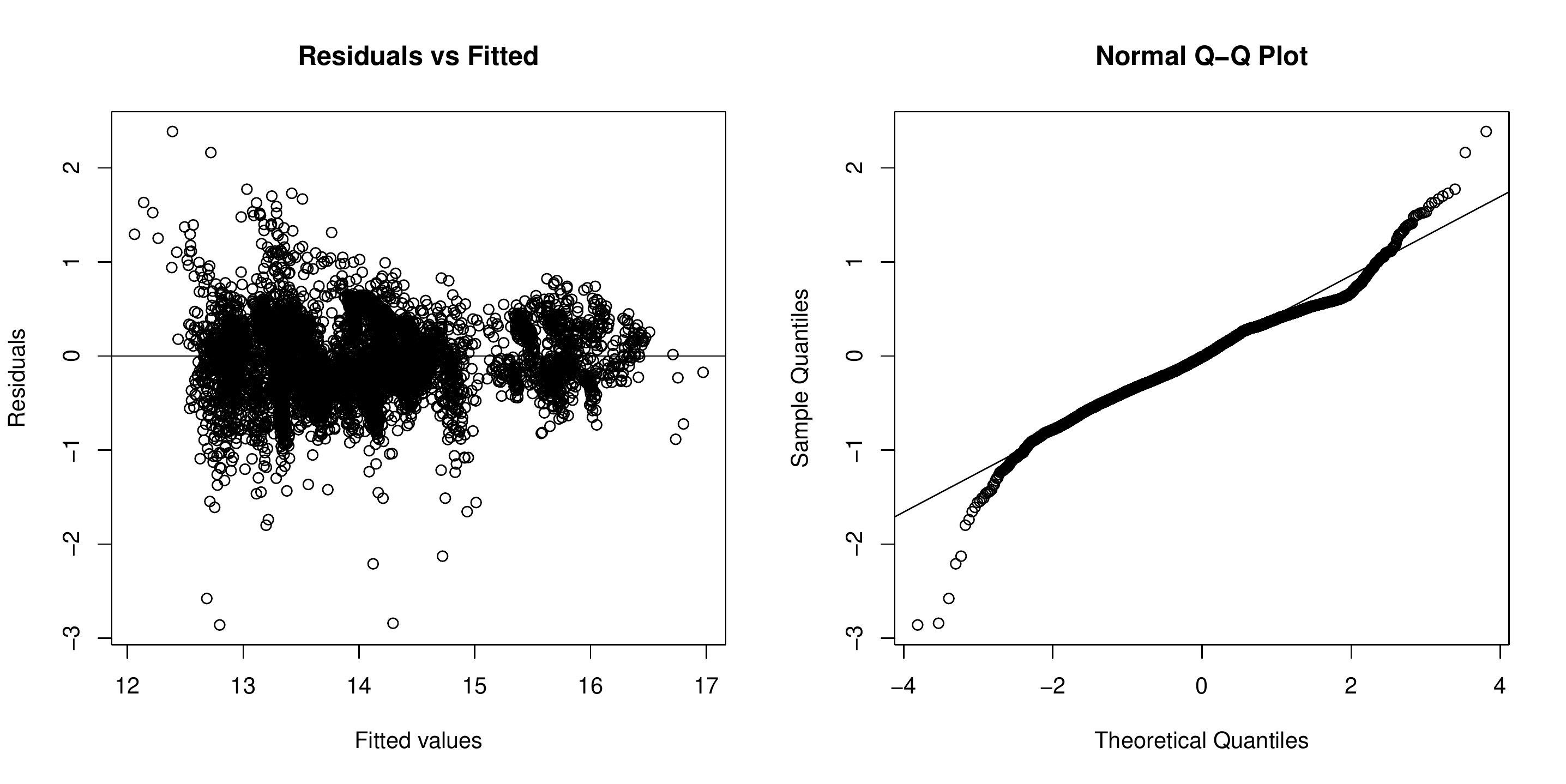} 
\par\end{centering}

\protect\protect\caption{\label{fig:model1}Regression model of $\log(y)$ against $\log(x)$
and strata indicator }
\end{figure}

The original population file contains $19,601$ retail businesses
stratified into $16$ strata, with a strata identifier ($h$), sales
($y$), and inventory values ($x$). For simulation purpose, we focus
on the first $5$ strata as a finite population, consisting of $7,260$
retail businesses. Figure \ref{fig:box plot} shows the scatter plot of
sales and inventory data by strata on a log scale.
%, where the data distribution differences by strata are obvious. 
We assumed the following superpopulation
model, 
\begin{equation}
\log(\text{y}_{hi})=\beta_{0h}+\beta_{1h}\log(x_{hi})+\epsilon_{hi},\label{eq:model1}
\end{equation}
%and 
%\begin{equation}
%\log(x_{i})=\alpha_{0h(i)}+\epsilon_{2i},\label{eq:model2}
%\end{equation}
where $\beta_{0h}$ and $\beta_{1h}$ are
strata-specific parameters with $h$ being the strata identifier, and $\epsilon_{hi}\sim N(0,\sigma_{h}^{2})$. To assess the adequacy
of model (\ref{eq:model1}), we made some diagnostic plots.
 Figure \ref{fig:model1}
shows the residual plot and the normal Q-Q plot for the fitted model
(\ref{eq:model1}). From the
residual plot, the constant variance assumption of $\epsilon_{hi}$
appears to be reasonable. From the normal Q-Q
plot, the normality assumption of $\epsilon_{hi}$ 
approximately holds.

To create missing, we considered univariate missing where only $y$
has missing values. We generated the response indicator $\delta$
of $y$ according to
\[
\delta\sim Bernoulli(\pi),\ \pi=1/[1+\exp\{4-0.3\log(x)\}].
\]
Under this model, the missing mechanism is MAR and the response rate
is about $0.6$. 

The parameters of interest are the stratum mean of $y$, $\eta_{h}=\mu_{h}$
for $1\leq h\leq5$, and the population mean of $y$, $\eta_{6}=\mu$.
The true parameter values are $\eta_{1}=92.25$,
$\eta_{2}=67.90$, $\eta_{3}=18.24$, $\eta_{4}=13.01$, $\eta_{5}=5.92$, and
$\eta_{6}=20.40$. The estimation methods included $(i)$ Full,
the full sample estimator, which is used as a benchmark for comparison,
$(ii)$ MI, the multiple imputation estimator with imputation size
$M=100$, and $(iii)$ PFI, the parametric fractional imputation estimator
with imputation size $M=100$, where the model parameters are estimated
by the pseudo MLE solving the score equation (4).

To generate samples, we considered stratified sampling with simple
random sampling within strata (STSRS) without replacement. Table \ref{tab:The-sample-allocation}
shows strata sizes $N_{h}$, sample sizes $n_{h}$, and sampling weights.
\textcolor{black}{The sampling weights range from $12.57$ to $45.79$.}\textcolor{red}{{}
}The samples are generated $2,000$ times.

\begin{table}[h]
\protect\protect\caption{\label{tab:The-sample-allocation}The sample allocation in stratified
simple random sampling. }

\centering{}%
\begin{tabular}{cccccc}
\hline 
Strata  & 1  & 2  & 3  & 4  & 5\tabularnewline
\hline 
Strata size $N_{h}$  & 352  & 566  & 1963  & 2181  & 2198\tabularnewline
Sample size $n_{h}$  & 28  & 32  & 46  & 46  & 48\tabularnewline
Sampling weight  & 12.57 & 17.69 & 42.67 & 47.41 & 45.79\tabularnewline
\hline 
\end{tabular}
\end{table}

For MI, we considered the imputation models  in (\ref{eq:model1}). Because the sampling design is stratified
random sampling and the imputation model includes the stratum indicator
function, the sampling design becomes noninformative. We first imputed
 $\log(y)$ from the posterior distribution of (\ref{eq:model1}), given the observed data, and then transformed
the imputations to the original scale of $y$. The implementation
of MI was carried out by the ``mice'' package in R. In each imputed
data set, we applied the following full-sample point estimators and
variance estimators: $\hat{\eta}_{1}=N^{-1}\sum_{h=1}^{H}N_{h}\bar{y}_{{h}}$
with $\bar{y}_{{h}}$ being the sample mean of $y$ in the $h$-th
stratum $S_{h}$, $\hat{V}(\hat{\eta})=N^{-2}\sum_{h=1}^{H}N_{h}^{2}(1-n_{h}/N_{h})s_{{h}}^{2}/n_{h}$
with $s_{{h}}^{2}=(n_{h}-1)^{-1}\sum_{i\in S_{h}}(y_{hi}-\bar{y}_{{h}})^{2}$. For
PFI, we considered the imputation model in (\ref{eq:model1}).\textcolor{black}{{} The proposal distribution in
the importance sampling step is the imputation distribution evaluated
at initial parameter values estimated from the available data. In
PFI, for estimating model parameters, we obtained the pseudo MLEs
by solving the score equations weighted by sampling
weights, as in (4). After imputation, $\eta$ was estimated by (5)
by choosing $U$ to be the corresponding estimating function. }We
used the delete-1 Jackknife replication method for variance estimation,
%as in \citet*{rao1992jackknife}, 
\[
\hat{V}_{R}(\hat{\eta})=\sum_{h=1}^{H}\frac{n_{h}-1}{n_{h}}\sum_{i\in S_{h}}(\hat{\eta}^{[i]}-\hat{\eta})^{2},
\]
where $\hat{\eta}^{[i]}$ is computed by omitting unit $i\in S_{h}$
and modifying the weights so that $w_{hj}$ is replaced by $n_{h}w_{hj}/(n_{h}-1)$
for all $j\in S_{h}$ and the weight remains the same for all other
$j$.

Table \ref{tab:Numercal-Results-of} shows the numerical results.
The mean and variance are calculated as the Monte Carlo mean and variance
of the point estimates across the simulated sample data. The relative
bias of the variance estimator is calculated as $\{(ve-var)/var\}$ $\times100\%$,
where $ve$ is the Monte Carlo mean of variance estimates and $var$ is
Monte Carlo variance of point estimates. In addition, $95\%$ confidence
intervals are calculated as $(\hat{\eta}-z_{0.975}\surd\hat{V},\hat{\eta}+z_{0.975}\surd\hat{V})$,
where $z_{0.975}$ is the $97.5\%$ quantile of the standard normal
distribution. The three estimators are essentially unbiased for point
estimation. The variances for PFI and MI are close for all parameters.
However, for inference, the validity of Rubin's variance estimator
relies on the congeniality condition (\citealt{meng1994multiple}),
which holds when MLEs are used as the full-sample estimator in MI,
but not for Method-of-Moments estimators (MMEs) 
under MAR (\citealt{yang2015mi}). As shown in Table \ref{tab:Numercal-Results-of},
Rubin's variance estimator of the MI estimator is biased upward for
strata means and the population mean with relative bias $48.06\%$,
$30.53\%$, $23.05\%$, $23.02\%$, $16.96\%$ for $\hat{\mu}_{j,MI}$,
$1\leq j\leq5$ and $32.75\%$ for $\hat{\mu}_{MI}$. Under the log
normal distribution and MAR, the MMEs are not self-efficient and Rubin's
variance estimator is biased, which is consistent with the results
in \citet*{meng1994multiple} and \citet*{yang2015mi}. Among those, Stratum 1 has largest
bias of the variance estimator, followed by Stratum 2, given their smaller sample sizes compared
to other strata. 
In addition, the mean width of confidence intervals is larger than that of FI.
For the population mean, we used the Horvitz\textendash Thompson
(HT) estimator as the full-sample estimator
instead of the MLE under log-normal distribution. It is well-known that
the HT estimator is robust but inefficient, which results in bias
in Rubin's variance estimator. The coverage of $95\%$ confidence interval reaches $98.3\%$ for
the population mean due to variance overestimation. In contrast, PFI
variance estimators applied to the HT estimator are essentially unbiased and provides
empirical coverages close to the nominal coverage. 

\begin{sidewaystable}
\protect\protect\caption{\label{tab:Numercal-Results-of}Numerical Results of Point Estimation
(Mean and Var), Relative Bias (R.B.) of Variance Estimation, Mean Width
and Coverage of $95\%$ Confidence Intervals (C.I.s) under Stratified Simple Random Sampling over $2,000$
Samples. The estimation methods include (i) FULL: the full sample
estimator, (ii) MI: the multiple imputation estimator with imputation
size $M=100$, (iii) PFI, the parametric fractional imputation estimator
with imputation size $M=100$, where the model parameters are obtained
by the pseudo MLE. The parameters are $\eta_{1}=$Stratum 1 mean,
$\eta_{2}=$Stratum 2 mean, $\eta_{3}=$Stratum 3 mean, $\eta_{4}=$Stratum
4 mean, $\eta_{5}=$Stratum 5 mean, $\eta_{6}=$Population mean.}

\centering{}%
\begin{tabular}{cccccccccccccccc}
\hline 
 & \multicolumn{3}{c}{Mean} & \multicolumn{3}{c}{Var} & \multicolumn{3}{c}{R.B. ($\%$)} & \multicolumn{3}{c}{Mean Width of C.I.s} & \multicolumn{3}{c}{Coverage}\tabularnewline
 & FULL & MI & PFI & FULL & MI & PFI & FULL & MI & PFI & FULL & MI & PFI & FULL & MI & PFI\tabularnewline
\hline 
$\eta_{1}$  & 92.46 & 93.95 & 92.85 & 76.46 & 119.18 & 120.67 & 6.08 & \textcolor{black}{48.06} & 7.81 & 18.01 & \textcolor{black}{26.57} & \textcolor{black}{22.81} & 0.951 & \textcolor{black}{0.964} & 0.952\tabularnewline
$\eta_{2}$  & 67.72 & 68.40 & 67.76 & 40.05 & 60.91 & 59.53 & 6.55 & \textcolor{black}{30.53} & 3.26 & 13.07  & \textcolor{black}{17.83} & \textcolor{black}{15.68} & 0.943 & \textcolor{black}{0.954} & 0.946\tabularnewline
$\eta_{3}$  & 18.30 & 18.45 & 18.28 & 2.12 & 3.32 & 3.29 & -3.06 & \textcolor{black}{23.05} & -1.63 & 2.86 & \textcolor{black}{4.04} & \textcolor{black}{3.60} & 0.944 & \textcolor{black}{0.961} & 0.948\tabularnewline
$\eta_{4}$  & 13.03 & 13.12 & 13.00 & 1.02 & 1.77 & 1.76 & 0.51 & \textcolor{black}{23.02} & -4.28 & 2.03 & \textcolor{black}{2.95} & \textcolor{black}{2.60} & 0.946 & \textcolor{black}{0.962} & 0.943\tabularnewline
$\eta_{5}$  & 5.92 & 5.98 & 5.91 & 0.22 & 0.46 & 0.46 & 1.84 & \textcolor{black}{16.96} & -4.40 & 0.94 & \textcolor{black}{1.47} & \textcolor{black}{1.32} & 0.953 & \textcolor{black}{0.963} & 0.947\tabularnewline
$\eta_{6}$  & 20.42 & 20.63 & 20.42 & 0.70 & 1.11 & 1.10 & -3.36 & \textcolor{black}{32.75} & -3.97 & 1.65 & \textcolor{black}{2.42} & \textcolor{black}{2.06} & 0.952 & \textcolor{black}{0.983} & 0.953\tabularnewline
\hline 
\end{tabular}
\end{sidewaystable}

\section{CONCLUDING REMARKS}

In survey sampling, MI and FI are two 
 available 
approaches of imputation for general-purpose estimation. In MI,   Rubin's
variance estimation formula is recommended because of its simplicity,
but it requires the congeniality condition of \citet*{meng1994multiple}, which can be restrictive  in practice.   A merit of FI is that the congeniality
condition is not needed for consistent variance estimation.
When the sampling design is informative, MI can use an augmented model to make the sampling design noninformative. However, incorporating all design information into the model is not always possible (\citealt{reiter2006importance}) and valid inference under MI  is not easy or sometimes impossible (\citealt{berg2015}).
In contrast, FI can handle informative sampling design easily as it incorporates sampling weights into estimation instead of modeling. 

So far, we have presented FI 
under the MAR case. Parametric FI can be adapted to a situation,
where the missing values are suspected to be missing not at random
(MNAR) (\citealt{kim2012parametric};
\citealt{Yang2013parametric}). A semiparametric FI using the exponential tilting model of \citet*{kimyu11} is also promising, which is under development. Also,
FI can be used
to approximate observed log likelihood easily (\citealt{Yang2015likelihood-based}). 
The approximation of the observed log likelihood can be directly applied
to model selections or model comparisons with missing data, such as
using Akaike Information Criterion or the Bayesian Information Criterion.
Further investigation on this topic will be worthwhile. 

We conclude the paper with the hope that \textcolor{black}{continuing
efforts will be made into developing statistical methods and corresponding
computational programs (an R software package is in progress) for
FI, so as to make these methods accessible to a broader audience.}

%\section*{Acknowledgements}
%And this is an acknowledgements section with a heading that was produced by the
%$\backslash$section* command. Thank you all for helping me writing this
%\LaTeX\ sample file.

\bibliography{pfi_kim}

\begin{thebibliography}{}

\bibitem[\protect\citeauthoryear{Andridge and Little}{Andridge and
  Little}{2010}]{andridge2010review}
Andridge, R.~R. and R.~J. Little (2010).
\newblock A review of hot deck imputation for survey non-response.
\newblock {\em Int. Stat. Rev.\/}~{\em 78\/}(1), 40--64.

\bibitem[\protect\citeauthoryear{Bang and Robins}{Bang and
  Robins}{2005}]{bang2005doubly}
Bang, H. and J.~M. Robins (2005).
\newblock Doubly robust estimation in missing data and causal inference models.
\newblock {\em Biometrics\/}~{\em 61\/}(4), 962--973.

\bibitem[\protect\citeauthoryear{Berg, Kim, and Skinner}{Berg
  et~al.}{2015}]{berg2015}
Berg, E., J.~K. Kim, and C.~Skinner (2015).
\newblock Imputation under informative sampling.
\newblock {\em Submitted\/}.

\bibitem[\protect\citeauthoryear{Binder and Patak}{Binder and
  Patak}{1994}]{binder1994use}
Binder, D.~A. and Z.~Patak (1994).
\newblock Use of estimating functions for estimation from complex surveys.
\newblock {\em J. Amer. Statist. Assoc.\/}~{\em 89\/}(427), 1035--1043.

\bibitem[\protect\citeauthoryear{Binder and Sun}{Binder and
  Sun}{1996}]{binder1996frequency}
Binder, D.~A. and W.~Sun (1996).
\newblock Frequency valid multiple imputation for surveys with a complex
  design.
\newblock In {\em Proceedings of the Survey Research Methods Section of the
  American Statistical Association}, pp.\  281--286.

\bibitem[\protect\citeauthoryear{Breidt and Fuller}{Breidt and
  Fuller}{1996}]{breidt96}
Breidt, F.~J., M.~A. and W.~A. Fuller (1996).
\newblock Two-phase sampling by imputation.
\newblock {\em Journal of Indian Society of Agricultural Statistics (Golden
  Jubilee Number)\/}~{\em 49}, 79--90.

\bibitem[\protect\citeauthoryear{Cao, Tsiatis, and Davidian}{Cao
  et~al.}{2009}]{cao2009improving}
Cao, W., A.~A. Tsiatis, and M.~Davidian (2009).
\newblock Improving efficiency and robustness of the doubly robust estimator
  for a population mean with incomplete data.
\newblock {\em Biometrika\/}~{\em 96\/}(3), 723--734.

\bibitem[\protect\citeauthoryear{Chen and Shao}{Chen and
  Shao}{2001}]{chen2001jackknife}
Chen, J. and J.~Shao (2001).
\newblock Jackknife variance estimation for nearest-neighbor imputation.
\newblock {\em J. Amer. Statist. Assoc.\/}~{\em 96\/}(453), 260--269.

\bibitem[\protect\citeauthoryear{Durrant}{Durrant}{2005}]{durrant2005imputation}
Durrant, G.~B. (2005).
\newblock Imputation methods for handling item-nonresponse in the social
  sciences: a methodological review.
\newblock {\em ESRC National Centre for Research Methods and Southampton Stat.
  Sci.s Research Institute. NCRM Methods Review Papers NCRM/002\/}.

\bibitem[\protect\citeauthoryear{Durrant}{Durrant}{2009}]{durrant2009imputation}
Durrant, G.~B. (2009).
\newblock Imputation methods for handling item-nonresponse in practice:
  methodological issues and recent debates.
\newblock {\em International Journal of Social Research Methodology\/}~{\em
  12\/}(4), 293--304.

\bibitem[\protect\citeauthoryear{Durrant and Skinner}{Durrant and
  Skinner}{2006}]{durrant2006using}
Durrant, G.~B. and C.~Skinner (2006).
\newblock Using missing data methods to correct for measurement error in a
  distribution function.
\newblock {\em Surv. Methodol.\/}~{\em 32\/}(1), 25.

\bibitem[\protect\citeauthoryear{Fay}{Fay}{1992}]{fay1992inferences}
Fay, R.~E. (1992).
\newblock When are inferences from multiple imputation valid?
\newblock In {\em Proceedings of the Survey Research Methods Section of the
  American Statistical Association}, Volume~81, pp.\  227--32.

\bibitem[\protect\citeauthoryear{Fay}{Fay}{1996}]{fay1996alternative}
Fay, R.~E. (1996).
\newblock Alternative paradigms for the analysis of imputed survey data.
\newblock {\em J. Amer. Statist. Assoc.\/}~{\em 91\/}(434), 490--498.

\bibitem[\protect\citeauthoryear{Fuller}{Fuller}{2003}]{fuller03}
Fuller, W.~A. (2003).
\newblock Estimation for multiple phase samples.
\newblock In R.~L. Chambers and C.~J. Skinner (Eds.), {\em Analysis of Survey
  Data}, pp.\  307--322. Chichester, U.~K.: John Wiley {\&} Sons.

\bibitem[\protect\citeauthoryear{Fuller and Kim}{Fuller and
  Kim}{2005}]{fuller2005hot}
Fuller, W.~A. and J.~K. Kim (2005).
\newblock Hot deck imputation for the response model.
\newblock {\em Surv. Methodol.\/}~{\em 31\/}(2), 139.

\bibitem[\protect\citeauthoryear{Godambe and Thompson}{Godambe and
  Thompson}{1986}]{godambe1986parameters}
Godambe, V. and M.~E. Thompson (1986).
\newblock Parameters of superpopulation and survey population: their
  relationships and estimation.
\newblock {\em Int. Stat. Rev./Revue Internationale de Statistique\/}~{\em
  54\/}(2), 127--138.

\bibitem[\protect\citeauthoryear{Hastings}{Hastings}{1970}]{hastings1970monte}
Hastings, W.~K. (1970).
\newblock {M}onte {C}arlo sampling methods using {M}arkov {C}hains and their
  applications.
\newblock {\em Biometrika\/}~{\em 57\/}(1), 97--109.

\bibitem[\protect\citeauthoryear{Haziza}{Haziza}{2009}]{haziza2009imputation}
Haziza, D. (2009).
\newblock Imputation and inference in the presence of missing data.
\newblock {\em Handbook of Statistics, Sample Surveys: Theory Methods and
  Inference, Editors: C.R. Rao and D. Pfeffermann\/}~{\em 29}, 215--246.

\bibitem[\protect\citeauthoryear{Ibrahim}{Ibrahim}{1990}]{Ibrahim90}
Ibrahim, J.~G. (1990).
\newblock Incomplete data in generalized linear models.
\newblock {\em J. Amer. Statist. Assoc.\/}~{\em 85}, 765--769.

\bibitem[\protect\citeauthoryear{Im, Kim, and Fuller}{Im et~al.}{2015}]{im2015}
Im, J., J.~K. Kim, and W.~A. Fuller (2015).
\newblock Two-phase sampling approach to fractional hot deck imputation.
\newblock {\em Submitted\/}.

\bibitem[\protect\citeauthoryear{Kalton and Kish}{Kalton and
  Kish}{1984}]{kalton1984some}
Kalton, G. and L.~Kish (1984).
\newblock Some efficient random imputation methods.
\newblock {\em Comm. Statist. Theory Methods.\/}~{\em 13\/}(16), 1919--1939.

\bibitem[\protect\citeauthoryear{Kang and Schafer}{Kang and
  Schafer}{2007}]{kang2007demystifying}
Kang, J.~D. and J.~L. Schafer (2007).
\newblock Demystifying double robustness: A comparison of alternative
  strategies for estimating a population mean from incomplete data.
\newblock {\em Stat. Sci.\/}~{\em 22\/}(4), 523--539.

\bibitem[\protect\citeauthoryear{Kim, Berg, and Park}{Kim
  et~al.}{2015}]{kimberg2015}
Kim, J., E.~Berg, and T.~Park (2015).
\newblock Statistical matching using fractional imputation.
\newblock Unpublished manuscript.

\bibitem[\protect\citeauthoryear{Kim}{Kim}{2011}]{kim11}
Kim, J.~K. (2011).
\newblock Parametric fractional imputation for missing data analysis.
\newblock {\em Biometrika\/}~{\em 98}, 119--132.

\bibitem[\protect\citeauthoryear{Kim, Brick, Fuller, and Kalton}{Kim
  et~al.}{2006}]{kim2006bias}
Kim, J.~K., J.~Brick, W.~A. Fuller, and G.~Kalton (2006).
\newblock On the bias of the multiple-imputation variance estimator in survey
  sampling.
\newblock {\em J. R. Stat. Soc. Ser. B. Stat. Methodol.\/}~{\em 68\/}(3),
  509--521.

\bibitem[\protect\citeauthoryear{Kim and Fuller}{Kim and
  Fuller}{2004}]{kim2004fractional}
Kim, J.~K. and W.~Fuller (2004).
\newblock Fractional hot deck imputation.
\newblock {\em Biometrika\/}~{\em 91\/}(3), 559--578.

\bibitem[\protect\citeauthoryear{Kim, Fuller, Bell, et~al.}{Kim
  et~al.}{2011}]{kim2011variance}
Kim, J.~K., W.~A. Fuller, W.~R. Bell, et~al. (2011).
\newblock Variance estimation for nearest neighbor imputation for {U.S.} census
  long form data.
\newblock {\em Ann. Appl. Stat.\/}~{\em 5\/}(2A), 824--842.

\bibitem[\protect\citeauthoryear{Kim and Haziza}{Kim and
  Haziza}{2014}]{kim2013doubly}
Kim, J.~K. and D.~Haziza (2014).
\newblock Doubly robust inference with missing data in survey sampling.
\newblock {\em Statist. Sinica\/}~{\em 24}, 375--394.

\bibitem[\protect\citeauthoryear{Kim and Hong}{Kim and Hong}{2012}]{kimhong12}
Kim, J.~K. and M.~Hong (2012).
\newblock An imputation approach to statistical inference with coarse data.
\newblock {\em Canad. J. Statist.\/}~{\em 40}, 604--618.

\bibitem[\protect\citeauthoryear{Kim, Navarro, and Fuller}{Kim
  et~al.}{2006}]{kim06b}
Kim, J.~K., A.~Navarro, and W.~A. Fuller (2006).
\newblock Replicate variance estimation after multi-phase stratified sampling.
\newblock {\em J. Amer. Statist. Assoc.\/}~{\em 101}, 312--320.

\bibitem[\protect\citeauthoryear{Kim and Rao}{Kim and Rao}{2012}]{kimrao12}
Kim, J.~K. and J.~N.~K. Rao (2012).
\newblock Combining data from two independent surveys: a model-assisted
  approach.
\newblock {\em Biometrika\/}~{\em 99}, 85--100.

\bibitem[\protect\citeauthoryear{Kim and Shao}{Kim and
  Shao}{2013}]{kim2013statistical}
Kim, J.~K. and J.~Shao (2013).
\newblock {\em Statistical Methods for Handling Incomplete Data}.
\newblock CRC Press.

\bibitem[\protect\citeauthoryear{Kim and Yang}{Kim and
  Yang}{2014}]{Kim2014Fractionalhotdeck}
Kim, J.~K. and S.~Yang (2014).
\newblock Fractional hot deck imputation for robust inference under item
  nonresponse in survey sampling.
\newblock {\em Surv. Methodol.\/}~{\em 40\/}(2), 211--230.

\bibitem[\protect\citeauthoryear{Kim and Yu}{Kim and Yu}{2011a}]{kimyu11b}
Kim, J.~K. and C.~L. Yu (2011a).
\newblock Replication variance estimation under two-phase sampling.
\newblock {\em Surv. Methodol.\/}~{\em 37}, 67--74.

\bibitem[\protect\citeauthoryear{Kim and Yu}{Kim and Yu}{2011b}]{kimyu11}
Kim, J.~K. and C.~L. Yu (2011b).
\newblock A semi-parametric estimation of mean functionals with non-ignorable
  missing data.
\newblock {\em J. Amer. Statist. Assoc.\/}~{\em 106}, 157--165.

\bibitem[\protect\citeauthoryear{Kim and Kim}{Kim and
  Kim}{2012}]{kim2012parametric}
Kim, J.~Y. and J.~K. Kim (2012).
\newblock Parametric fractional imputation for nonignorable missing data.
\newblock {\em J. Korean Statist. Soc.\/}~{\em 41\/}(3), 291--303.

\bibitem[\protect\citeauthoryear{Kitamura, Tripathi, and Ahn}{Kitamura
  et~al.}{2009}]{kitamura2009variance}
Kitamura, Y., G.~Tripathi, and H.~Ahn (2009).
\newblock Variance estimation when donor imputation is used to fill in missing
  values.
\newblock {\em Canad. J. Statist.\/}~{\em 37\/}(3), 400--416.

\bibitem[\protect\citeauthoryear{Kott}{Kott}{1995}]{kott1995paradox}
Kott, P. (1995).
\newblock A paradox of multiple imputation.
\newblock In {\em Proceedings of the Survey Research Methods Section of the
  American Statistical Association}, pp.\  384--389.

\bibitem[\protect\citeauthoryear{Little and Rubin}{Little and
  Rubin}{2002}]{little2002statistical}
Little, R.~J. and D.~B. Rubin (2002).
\newblock Statistical analysis with missing data.

\bibitem[\protect\citeauthoryear{Louis}{Louis}{1982}]{louis82}
Louis, T.~A. (1982).
\newblock Finding the observed information matrix when using the {EM}
  algorithm.
\newblock {\em J. R. Stat. Soc. Ser. B. Stat. Methodol.\/}~{\em 44\/}(2),
  226--233.

\bibitem[\protect\citeauthoryear{Meng}{Meng}{1994}]{meng1994multiple}
Meng, X.-L. (1994).
\newblock Multiple-imputation inferences with uncongenial sources of input.
\newblock {\em Stat. Sci.\/}~{\em 9}, 538--558.

\bibitem[\protect\citeauthoryear{Mulry, Oliver, and Kaputa}{Mulry
  et~al.}{2014}]{mulry2014detecting}
Mulry, M.~H., B.~E. Oliver, and S.~J. Kaputa (2014).
\newblock Detecting and treating verified influential values in a monthly
  retail trade survey.
\newblock {\em Journal of Official Statistics\/}~{\em 30\/}(4), 721--747.

\bibitem[\protect\citeauthoryear{Nadaraya}{Nadaraya}{1964}]{nadaraya1964estimating}
Nadaraya, E.~A. (1964).
\newblock On estimating regression.
\newblock {\em Theory Probab. Appl.\/}~{\em 9\/}(1), 141--142.

\bibitem[\protect\citeauthoryear{Nielsen}{Nielsen}{2003}]{nielsen2003proper}
Nielsen, S.~F. (2003).
\newblock Proper and improper multiple imputation.
\newblock {\em Int. Stat. Rev.\/}~{\em 71\/}(3), 593--607.

\bibitem[\protect\citeauthoryear{Pfeffermann, Skinner, Holmes, Goldstein, and
  Rasbash}{Pfeffermann et~al.}{1998}]{pfeffermann1998weighting}
Pfeffermann, D., C.~J. Skinner, D.~J. Holmes, H.~Goldstein, and J.~Rasbash
  (1998).
\newblock Weighting for unequal selection probabilities in multilevel models.
\newblock {\em J. R. Stat. Soc. Ser. B. Stat. Methodol.\/}~{\em 60\/}(1),
  23--40.

\bibitem[\protect\citeauthoryear{Rao}{Rao}{1973}]{rao1973double}
Rao, J. (1973).
\newblock On double sampling for stratification and analytical surveys.
\newblock {\em Biometrika\/}~{\em 60\/}(1), 125--133.

\bibitem[\protect\citeauthoryear{Rao, Yung, and Hidiroglou}{Rao
  et~al.}{2002}]{rao2002estimating}
Rao, J., W.~Yung, and M.~Hidiroglou (2002).
\newblock Estimating equations for the analysis of survey data using
  poststratification information.
\newblock {\em Sankhy{\=a}: The Indian Journal of Statistics, Series A\/}~{\em
  64\/}(2), 364--378.

\bibitem[\protect\citeauthoryear{Rao and Shao}{Rao and
  Shao}{1992}]{rao1992jackknife}
Rao, J.~N. and J.~Shao (1992).
\newblock Jackknife variance estimation with survey data under hot deck
  imputation.
\newblock {\em Biometrika\/}~{\em 79\/}(4), 811--822.

\bibitem[\protect\citeauthoryear{Reiter, Raghunathan, and Kinney}{Reiter
  et~al.}{2006}]{reiter2006importance}
Reiter, J.~P., T.~E. Raghunathan, and S.~K. Kinney (2006).
\newblock The importance of modeling the sampling design in multiple imputation
  for missing data.
\newblock {\em Surv. Methodol.\/}~{\em 32\/}(2), 143.

\bibitem[\protect\citeauthoryear{Robins, Rotnitzky, and Zhao}{Robins
  et~al.}{1994}]{robins1994estimation}
Robins, J.~M., A.~Rotnitzky, and L.~P. Zhao (1994).
\newblock Estimation of regression coefficients when some regressors are not
  always observed.
\newblock {\em J. Amer. Statist. Assoc.\/}~{\em 89\/}(427), 846--866.

\bibitem[\protect\citeauthoryear{Robins and Wang}{Robins and
  Wang}{2000}]{robins2000inference}
Robins, J.~M. and N.~Wang (2000).
\newblock Inference for imputation estimators.
\newblock {\em Biometrika\/}~{\em 87\/}(1), 113--124.

\bibitem[\protect\citeauthoryear{Rubin}{Rubin}{1976}]{rubin1976inference}
Rubin, D.~B. (1976).
\newblock Inference and missing data.
\newblock {\em Biometrika\/}~{\em 63\/}(3), 581--592.

\bibitem[\protect\citeauthoryear{Rubin}{Rubin}{1987}]{rubin1987multiple}
Rubin, D.~B. (1987).
\newblock {\em Multiple Imputation for Nonresponse in Surveys}.
\newblock John Wiley \& Sons.

\bibitem[\protect\citeauthoryear{Rubin}{Rubin}{1996}]{rubin1996multiple}
Rubin, D.~B. (1996).
\newblock Multiple imputation after 18+ years.
\newblock {\em J. Amer. Statist. Assoc.\/}~{\em 91\/}(434), 473--489.

\bibitem[\protect\citeauthoryear{{SAS Institute Inc.}}{{SAS Institute
  Inc.}}{2015}]{SAS2015}
{SAS Institute Inc.} (2015).
\newblock {SAS/STAT 14.1 User's Guide} - the {SURVEYIMPUTE} procedure, {Cary,
  NC: SAS Institue Inc.}

\bibitem[\protect\citeauthoryear{Schafer}{Schafer}{1997}]{schafer1997imputation}
Schafer, J.~L. (1997).
\newblock Imputation of missing covariates under a multivariate linear mixed
  model.
\newblock {\em Unpublished technical report\/}.

\bibitem[\protect\citeauthoryear{Schenker and Raghunathan}{Schenker and
  Raghunathan}{2007}]{schenker07}
Schenker, N. and T.~Raghunathan (2007).
\newblock Combining information from multiple surveys to enhance estimation of
  measures of health.
\newblock {\em Statist. Med.\/}~{\em 26}, 1802--11.

\bibitem[\protect\citeauthoryear{Schenker, Raghunathan, Chiu, Makuc, Zhang, and
  Cohen}{Schenker et~al.}{2006}]{schenker2006multiple}
Schenker, N., T.~E. Raghunathan, P.-L. Chiu, D.~M. Makuc, G.~Zhang, and A.~J.
  Cohen (2006).
\newblock Multiple imputation of missing income data in the national health
  interview survey.
\newblock {\em J. Amer. Statist. Assoc.\/}~{\em 101\/}(475), 924--933.

\bibitem[\protect\citeauthoryear{Smith and Gelfand}{Smith and
  Gelfand}{1992}]{smith92}
Smith, A.~F.~M. and A.~E. Gelfand (1992).
\newblock Bayesian statistics without tears: A sampling-resampling perspective.
\newblock {\em Amer. Statist.\/}~{\em 46}, 84--88.

\bibitem[\protect\citeauthoryear{Tan}{Tan}{2006}]{tan2006distributional}
Tan, Z. (2006).
\newblock A distributional approach for causal inference using propensity
  scores.
\newblock {\em J. Amer. Statist. Assoc.\/}~{\em 101\/}(476), 1619--1637.

\bibitem[\protect\citeauthoryear{Tanner and Wong}{Tanner and
  Wong}{1987}]{Tanner87}
Tanner, M.~A. and W.~H. Wong (1987).
\newblock The calculation of posterior distributions by data augmentation (with
  discussion).
\newblock {\em J. Amer. Statist. Assoc.\/}~{\em 82\/}(398), 528--550.

\bibitem[\protect\citeauthoryear{Vink, Frank, Pannekoek, and Buuren}{Vink
  et~al.}{2014}]{vink2014predictive}
Vink, G., L.~E. Frank, J.~Pannekoek, and S.~Buuren (2014).
\newblock Predictive mean matching imputation of semicontinuous variables.
\newblock {\em Stat. Neerl.\/}~{\em 68\/}(1), 61--90.

\bibitem[\protect\citeauthoryear{Wang and Chen}{Wang and
  Chen}{2009}]{wang2009empirical}
Wang, D. and S.~X. Chen (2009).
\newblock Empirical likelihood for estimating equations with missing values.
\newblock {\em Ann. Statist.\/}~{\em 37\/}(1), 490--517.

\bibitem[\protect\citeauthoryear{Wang and Robins}{Wang and
  Robins}{1998}]{wang1998large}
Wang, N. and J.~M. Robins (1998).
\newblock Large-sample theory for parametric multiple imputation procedures.
\newblock {\em Biometrika\/}~{\em 85\/}(4), 935--948.

\bibitem[\protect\citeauthoryear{Wei and Tanner}{Wei and
  Tanner}{1990}]{wei1990monte}
Wei, G.~C. and M.~A. Tanner (1990).
\newblock A {M}onte {C}arlo implementation of the {EM} algorithm and the poor
  man's data augmentation algorithms.
\newblock {\em J. Amer. Statist. Assoc.\/}~{\em 85\/}(411), 699--704.

\bibitem[\protect\citeauthoryear{Yang and Kim}{Yang and
  Kim}{2014}]{Yang2014SemiparametricInference}
Yang, S. and J.~K. Kim (2014).
\newblock A semiparametric inference to regression analysis with missing
  covariates in survey data.
\newblock {\em Submitted\/}.

\bibitem[\protect\citeauthoryear{Yang and Kim}{Yang and
  Kim}{2015a}]{Yang2015likelihood-based}
Yang, S. and J.~K. Kim (2015a).
\newblock Likelihood-based inference with missing data under missing-at-random.
\newblock {\em Scand. J. Stat.\/}, accepted.

\bibitem[\protect\citeauthoryear{Yang and Kim}{Yang and
  Kim}{2015b}]{yang2015mi}
Yang, S. and J.~K. Kim (2015b).
\newblock A note on multiple imputation for method of moments estimation.
\newblock {\em Submitted\/}.

\bibitem[\protect\citeauthoryear{Yang, Kim, and Zhu}{Yang
  et~al.}{2013}]{Yang2013parametric}
Yang, S., J.~K. Kim, and Z.~Zhu (2013).
\newblock Parametric fractional imputation for mixed models with nonignorable
  missing data.
\newblock {\em Stat. Interface\/}~{\em 6}, 339--347.

\end{thebibliography}

\end{document}